\documentclass{article}

\usepackage{PRIMEarxiv}

\usepackage[utf8]{inputenc} 
\usepackage[T1]{fontenc}    
\usepackage{hyperref}       
\usepackage{url}            
\usepackage{booktabs}       
\usepackage{amsfonts,amsmath}       
\usepackage{cleveref}
\usepackage{nicefrac}       
\usepackage{microtype}      
\usepackage{lipsum}
\usepackage{fancyhdr}       
\usepackage{graphicx}       
\graphicspath{{media/}}     
\newtheorem{proposition}{Proposition}
\usepackage{natbib}
\usepackage[title]{appendix}
\usepackage[symbol]{footmisc}

\usepackage{tikz}
\usetikzlibrary{arrows.meta, positioning}

\title{Penalizing complexity priors for Bayesian inference of circular models
}

\author{
  Xiang Ye\textsuperscript{\dag}, Janet Van Niekerk\textsuperscript{*,\dag,\ddag}, Håvard Rue\textsuperscript{\dag} \\
  \vspace{5mm} \\ 
  \textsuperscript{\dag} Statistics Program \\
  King Abdullah University of Science and Technology \\
  Thuwal, Saudi Arabia \\
  \texttt{\{xiang.ye, haavard.rue\}@kaust.edu.sa} \\
  \vspace{3mm} \\
  \textsuperscript{\ddag} Department of Statistics \\
  University of Pretoria \\
  Pretoria, South Africa \\
  \texttt{janet.vanniekerk@up.ac.za}
}

\begin{document}
\maketitle

\begin{abstract}
Advancements in computational power and methodologies have enabled research on massive datasets. However, tools for analyzing data with directional or periodic characteristics, such as wind directions and customers' arrival time in 24-hour clock, remain underdeveloped. While statisticians have proposed circular distributions for such analyses, significant challenges persist in constructing circular statistical models, particularly in the context of Bayesian methods. These challenges stem from limited theoretical development and a lack of historical studies on prior selection for circular distribution parameters.

In this article, we propose a framework for selecting hyperpriors that contracts to a simpler model in circular scenarios, especially when there is insufficient information to guide prior selection. We introduce well-examined Penalized Complexity (PC) priors for the most widely used circular distributions. Comprehensive comparisons with existing hyperpriors in the literature are conducted through simulation studies and a practical case study. Finally, we discuss the contributions and implications of our work, providing a foundation for further advancements in constructing Bayesian circular statistical models.
\end{abstract}


\section{Introduction} \label{sec:introduction}

Advancements in computational power have provided researchers with access to massive and diverse datasets across various fields, enabling the exploration of complex scientific and practical challenges. However, tools for analyzing data with complex structures, such as directional or periodic characteristics, remain underdeveloped, and many of these datasets require specialized methods for analysis. For instance, wind directions and bird migration paths are naturally represented on a compass, while hospital patient arrival times and passenger density fluctuations follow a 24-hour clock \citep{mardia2009directional}. The orientation of earthquake epicenters, directions of cosmic rays and stellar objects in astronomy \citep{ley2017modern, pewsey2021recent,cabella2009statistical}, joint angles prone to injury, protein structure with angular measures in bioinformatics \citep{mardia2018directional, boomsma2008generative}, typhoon trajectories, and the analysis of angular components in multivariate extreme value statistics are additional examples of data that align with circular scales. These examples underscore the growing need for statistical frameworks capable of handling circular data.

Directional statistics offers appropriate tools for analyzing such data, with circular distributions — probability distributions defined on the circumference of a circle \citep{jammalamadaka2001topics} and characterized by angular measures or radians — playing a central role. These methods are crucial for modeling data in fields such as meteorology, earth sciences, bioinformatics, ecology, medicine \citep{vuollo2016analyzing, pardo2016directional}, genetics, neurology, astronomy \citep{cabella2009statistical, marinucci2011random}, image analysis \citep{jung2011principal, esteves2018learning}, text mining \citep{dhillon2001concept, banerjee2005clustering}, machine learning \citep{sra2016directional}, and beyond \citep{ley2017modern, pewsey2021recent}.

Applying Bayesian methods to circular data, however, poses significant challenges, particularly in the selection of priors — a pivotal step in Bayesian analysis. Unlike Euclidean distributions, circular distributions have unique parameterizations and behaviors, often requiring specialized approaches. For example, in \citet{wallace1993mml}, priors for the concentration parameter $\kappa$ of von Mises distribution are constructed through the techniques of Minimum Message Length (\Cref{sec:review_on_circular_distributions_and_priors.von_mises_distribution}). These priors are also used in \citet{dowe1996bayesian} and \citet{marrelec2024estimating}. Another popular prior for von Mises distribution is the joint conjugate prior proposed in \citet{damien1999full}. In addition, in the general procedure for Bayesian analysis with wrapped distributions proposed by \citet{ravindran2011bayesian}, the $\operatorname{Beta}\left(a,a\right)$ prior is employed. \citet{nunez2011bayesian} proposed to fit the Bayesian circular model through the projected normal distribution with normal priors. A prior for the location parameter is often more intuitive to formulate, while priors for the hyperparameters like dispersion parameters are notoriously hard to conceptualize. Moreover, when specific values of the hyperparamaters result in model complexity reduction, care should be taken in the prior construction.

Since circular variables are defined on a compact and curved space rather than in linear Euclidean space, their geometric properties make it difficult to develop strong intuition about their behavior. This further complicates prior selection, increasing the risk of using priors that either dominate the posterior, or possibly lead to poorly performing models. 
When a model can be viewed as a complex model containing a simpler counterpart, then a prior allocating insufficient mass to the \textit{simpler model} can result in inferring a complex model that is not supported by the data. A detailed discussion of the simpler model is presented in Section~\ref{sec:a_principled_framework_for_prior_selection.model_complexity_check}.

To address these challenges, we set two goals for this research. Firstly, we establish a prior selection framework for the hyperparameters of a general circular model, inspired by the Penalized Complexity (PC) prior framework proposed by \citet{simpson2017penalising}. The PC prior framework is a reliable choice for constructing default priors, as it balances model complexity and prior informativeness, particularly in scenarios with limited prior knowledge, where objective or uninformative priors are often considered. Second, we derive the explicit expressions for the PC priors for the most commonly used circular distributions' hyperparameters and provide a way to quantify prior information through a user-defined parameter. 

The structure of this paper is as follows: \Cref{sec:review_on_circular_distributions_and_priors} reviews the most commonly used circular distributions, their properties, and existing priors in the literature. \Cref{sec:a_principled_framework_for_prior_selection} introduces the proposed framework for prior selection, the procedure for deriving PC priors for circular distributions, and the formulations for widely used circular distributions. \Cref{sec:comparison_and_investigation_of_common_priors} evaluates the proposed priors through comparison studies and simulation studies, while \Cref{sec:application} demonstrates their application to real-world datasets. Finally, \Cref{sec:conclusions_and_outlook} discusses broader implications and potential directions for future research.

\section{Preliminaries} \label{sec:review_on_circular_distributions_and_priors}
A circular distribution should have a probability density function $p\left(x \mid \xi\right)$ that satisfies $p\left(x \mid \boldsymbol{\xi}\right) = p\left(x + 2\pi k \mid \boldsymbol{\xi}\right)$ for any integer $k$ and $x \in [0, 2\pi)$, where $\boldsymbol{\xi}$ is the parameters for the probability density function. Most circular distributions are constructed by using one of the following four general approaches: wrapping, conditioning, projection, and perturbation \citep{ley2017modern}.
  
Circular distributions in the wrapped family are constructed by taking a distribution defined on $\mathbb{R}$ and wrapping it around a circle (modulus by $2\pi$, e.g. wrapped Normal distribution, wrapped Cauchy distribution, wrapped double exponential distribution); the conditioning approach obtains circular distributions through constructing joint distribution of polar coordinates (radius $r$ and angle $\theta$) and finding the conditional distribution $p\left(\theta \mid r\right)$ through restricting the radius $r=1$ (e.g. von Mises distribution); the projection approach projects an distribution on $\mathbb{R}^{2}$ onto the unit circle (e.g. projected Normal distribution); The perturbation approach provides flexible choices to extend a circular density to a more general form through multiplying it with a proper function (e.g. cardioid distribution). Details can be found in Chapter 2.2 of \citet{ley2017modern}.
  
Notably, many of these distributions include the circular uniform distribution as a special case (\autoref{fig:relationship_between_popular_circular_distributions}), defined by a probability density function given by $p_{\mathcal{U}}\left(x\right) = \frac{1}{2\pi}, \quad x\in\left[0,2\pi\right)$.
\begin{figure}[ht]
\centering
\begin{tikzpicture}[
    >=Stealth,
    on grid,
    auto
  ]

  \node[draw, rectangle] (vonmises) {Von Mises};
  \node[draw, rectangle, right=4cm of vonmises] (circunif) {Circular Uniform};
  \node[draw, rectangle, right=4cm of circunif] (cardioid) {Cardioid};

  \node[draw, rectangle, below=1.8cm of vonmises] (pointmass) {Point Mass};
  \node[draw, rectangle, below=1.8cm of circunif] (wrapcauchy) {Wrapped Cauchy};
  \node[draw, rectangle, below=1.8cm of cardioid] (cardcurve) {Cardioid Curve};

  \draw[->] (vonmises) -- node[above] {$\kappa = 0$} (circunif);
  \draw[->] (vonmises) -- node[left]  {$\kappa \to \infty$} (pointmass);

  \draw[<-] (circunif) -- node[above] {$\ell = 0$} (cardioid);
  \draw[->] (cardioid) -- node[right] {$\ell \to \tfrac12$} (cardcurve);

  \draw[->] (wrapcauchy) -- (pointmass)
   node[midway, above, sloped]{\(\rho \to 1\)};
  \draw[->] (wrapcauchy) -- (circunif)
   node[midway, anchor=west, xshift=3pt]{\(\rho = 0\)};

\end{tikzpicture}
\caption{Relationship between popular circular distributions. $\kappa$, $\ell$ and $\rho$ are the concentration parameters for von Mises, cardioid and wrapped Cauchy distributions.}
\label{fig:relationship_between_popular_circular_distributions}
\end{figure}
Three of the most widely used circular distributions — the von Mises (vM) distribution, the cardioid distribution, and the wrapped Cauchy (WC) distribution — serve as the foundation for many generalizations and extensions in circular statistics \citep{ley2017modern}, such as the Jones-Pewsey distribution \citep{jones2005family} and the Kato-Jones distribution \citep{kato2010family}, underscoring their importance in the field. One commonality between these three distributions is that they all include two parameters: one location parameter and one concentration parameter. The concentration parameter is essentially a scaling parameter for circular distributions. When data is distributed on a unit circle (or sphere/hypersphere), the concept of the standard deviation, as defined for linear data, becomes meaningless, since its interpretation is no longer clear, particularly when the standard deviation exceeds half a circle ($\pi$). To describe the spread or sparsity of circular data, most circular distributions introduce a "concentration" parameter, which quantifies the extent of concentration (or dispersion) within the data, indicating how densely the data are clustered around a mean direction (or how widely they are spread). In this work, we focus on the von Mises (vM) distribution, the cardioid distribution, and the wrapped Cauchy (WC) distribution, and on their Bayesian inference.

\subsection{Von Mises Distribution} \label{sec:review_on_circular_distributions_and_priors.von_mises_distribution}

\begin{figure}[!ht]
    \centering
    \includegraphics[width=\linewidth]{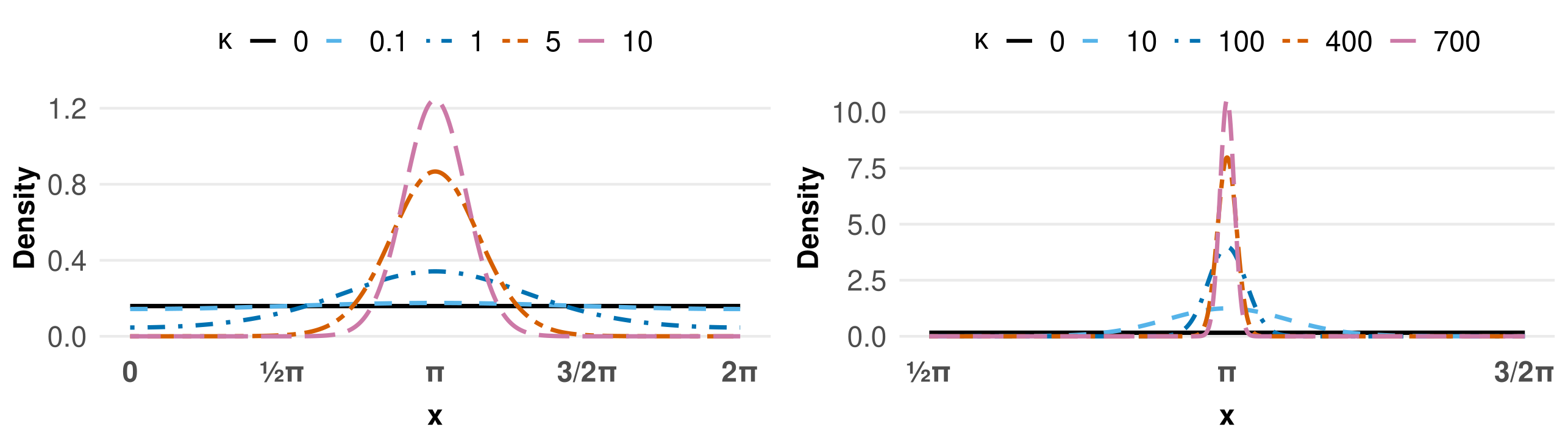}
    \caption{vM density for small (left) and large (right) $\kappa$ values with $\mu = \pi$ .}
    \label{fig:density_vm}
\end{figure}

The \textit{von Mises distribution} \citep{mardia2009directional}, often referred to as the \textit{circular normal distribution} \citep{gumbel1953circular}, is the circular analogue of the normal distribution. It is one of the most widely used and versatile circular distributions \citep{mardia2009directional}. Its probability density function is given by:
\begin{equation}
    \label{eq:von_Mises_density}
    \begin{aligned}
        p_{\mathcal{VM}}\left(x \mid \mu, \kappa\right) &= \frac{1}{2\pi \mathcal{I}_{0} \left(\kappa\right)} \exp \left\{ \kappa \cos \left( x - \mu \right) \right\}, \quad \mu \in \left[0,2\pi\right), \quad \kappa \in \left[0,\infty\right),
    \end{aligned}
\end{equation}
where $\mu$ is the location parameter, $\mathcal{I}_{a}\left(\cdot\right)$ is the modified Bessel function of the first kind of order $a \in \mathbb{N}$, and $\kappa$ is the concentration parameter. The plot for this density is given in \autoref{fig:density_vm}.
The von Mises distribution includes the circular uniform distribution as a special case when $\kappa = 0$, while $\kappa \to \infty$ indicates that the data are highly concentrated around the mean direction $\mu$. Thus, $\kappa$ can be interpreted as analogous to the precision of the normal distribution ($1/\sigma^{2}$). Since the Gamma distribution is a common prior choice for the precision in normal distributions, the $\operatorname{Gamma}\left(a, b\right)$ distribution with density given in \autoref{eq:gamma.density} is naively considered as a prior for $\kappa$.

\begin{equation}
    \label{eq:gamma.density}
    \begin{aligned}
        p\left(x \mid a,b\right) &= \frac{b^{a}}{\Gamma\left(a\right)}x^{a-1}\exp\left\{-bx\right\}
    \end{aligned}
\end{equation}

\citet{guttorp1988finding} proposed a joint conjugate prior for $\mu$ and $\kappa$, expressed as:
\begin{equation}
    \label{eq:von_mises_joint_conjugate_prior}
    \begin{aligned}
        p\left(\mu,\kappa\right) &\propto \left[ \mathcal{I}_{0} \left(\kappa\right) \right]^{-c} \exp \left\{ \kappa R_{0} \cos \left( x - \mu_{0} \right) \right\}
    \end{aligned}
\end{equation}
where $c$, $R_{0}$ and $\mu_{0}$ are prior hyperparameters. When $c \in \mathbb{N}$, $c$ can be interpreted as representing $c$ prior observations concentrated around the direction $\mu_{0}$, with $R_{0}$ corresponding to the component of the resultant vector in the known direction \citep{damien1999full}.

Despite its intuitive interpretation, deriving the posterior distribution under this prior requires introducing additional hyperparameters and several latent variables during the sampling process (see \citealt{damien1999full} for details). This construction, however, increases computational complexity and often results in greater Monte Carlo variability of the posterior estimates, owing to the mixing and dependence among latent components. In some cases, particularly when $c$ and $R_{0}$ are not well chosen, or when their combinations convey weak prior information, it may also lead to wider posterior credible intervals.

\citet{dowe1996bayesian} and \citet{marrelec2024estimating} both considered two different priors for $\kappa$ originally proposed by \citet{wallace1993mml} derived from minimum message length (MML). These priors are defined as:
\begin{equation}
    \label{eq:von_mises_kappa_h2_and_h3_prior}
    \begin{aligned}
        h_{2}\left(\kappa\right) = \frac{2}{\pi\left(1+\kappa^{2}\right)} \quad \text{and} \quad h_{3}\left(\kappa\right) = \frac{\kappa}{\left(1+\kappa^{2}\right)^{3/2}}.
    \end{aligned}
\end{equation}
Minimum Message Length (MML) is a Bayesian information-theoretic restatement of Occam’s razor, favoring models that describe the data most efficiently. It encodes both the model and the data as one message, where a more complex model requires a longer description and is only preferred if it leads to a shorter overall message. In this sense, priors derived from MML, such as $h_{2}\left(\kappa\right)$ and $h_{3}\left(\kappa\right)$, naturally penalize overly complex or highly concentrated models unless the data provide strong evidence to support them.

\subsection{Cardioid Distribution} \label{sec:review_on_circular_distributions_and_priors.cardioid_distribution}
\begin{figure}[!ht]
    \centering
    \includegraphics[width=\linewidth]{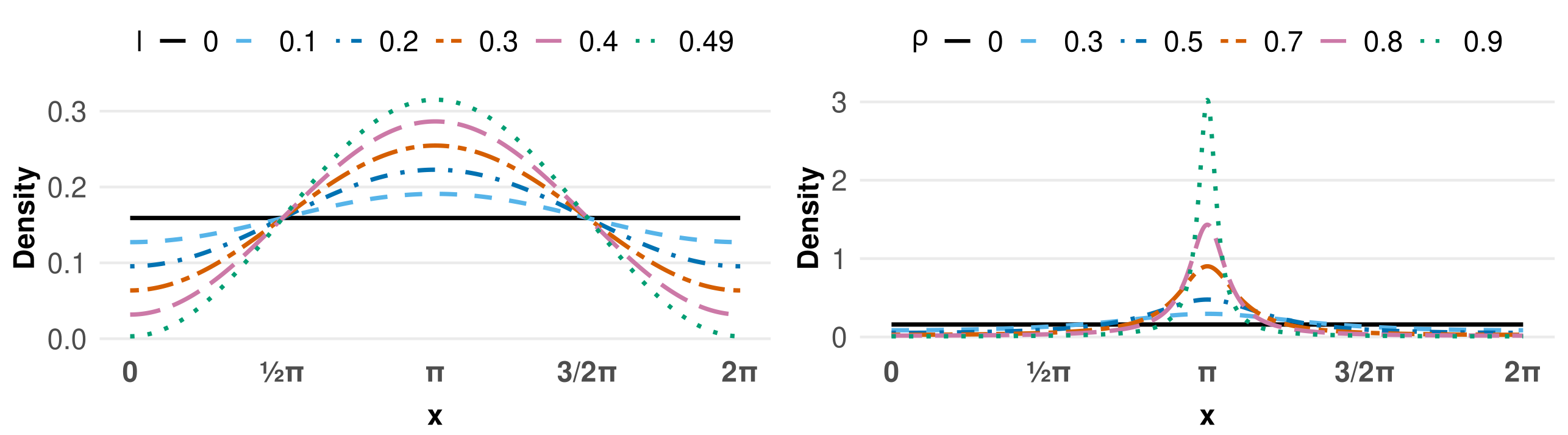}
    \caption{Cardioid density (left) and wrapped Cauchy density (right) with $\mu = \pi$ for different concentration parameter values.}
    \label{fig:density_card_and_wc}
\end{figure}

The \textit{cardioid distribution} a cardioid perturbation of the circular uniform distribution \citep{ley2017modern} with probability density function given by
\begin{equation}
    \label{eq:cardioid_density}
    \begin{aligned}
        p_{\mathcal{C}}\left(x \mid \mu, \ell\right) &= \frac{1}{2\pi}\left( 1+2\ell \cos\left(x-\mu\right) \right) , \quad \mu \in \left[0, 2\pi \right), \quad \ell \in \left[0,1/2\right).
    \end{aligned}
\end{equation}
Here, $\ell$ is the concentration parameter which controls the deviation from the circular uniform distribution. Larger values of $\ell$ result in stronger departures from uniformity (see the left-hand-side plot in \autoref{fig:density_card_and_wc}). It is obvious from the plot that the cardioid density is much less concentrated compared with the von Mises density, even when the concentration parameter $\ell$ for the former density approaches its supremum. This property enable the cardioid distribution to be more appropriate for dispersed data (heavy-tailed data) compared with von Mises distribution. Thus, the cardioid distribution is primarily used as a small concentration approximation to the von Mises distribution, as noted by \citet{mardia2009directional}.
Two reasonable prior choices for $\ell$ could be the $\operatorname{Uniform}\left(0,0.5\right)$ prior and the $0.5 \times \operatorname{Beta}\left(a,b\right)$ prior, since both priors fit the requirement of support being $\left[0,0.5\right)$.

\subsection{Wrapped Cauchy Distribution} \label{sec:review_on_circular_distributions_and_priors.wrapped_cauchy_distribution}

The wrapped Cauchy distribution is one of the rare wrapped distribution that has an analytic density function, since its probability density can be expressed in closed form \citep{ley2017modern}. Further details of the WC distribution are discussed in Chapter 3.5.7 of \citet{mardia2009directional}. The probability density function is given by:
\begin{equation}
    \label{eq:wrapped_Cauchy_density}
    \begin{aligned}
        p_{\mathcal{WC}}\left(x \mid \mu,\rho\right) = \frac{1}{2\pi} \frac{1 - \rho^2}{1 + \rho^2 - 2\rho \cos\left(x - \mu\right)}, \quad \text{where} \quad \mu \in [0,2\pi), \quad \rho \in \left[0,1\right).
    \end{aligned}
\end{equation}
The parameter $\rho$ serves as a measure of concentration, with $\rho = 0$ corresponding to the circular uniform distribution, indicating no concentration and as $\rho \to 1$, the data become increasingly concentrated, converging to a point mass distribution. The density plot for different $\rho$ values is given in the right-hand side of \autoref{fig:density_card_and_wc}.

For Bayesian analysis with a WC distribution, one reasonable choice for the prior for $\rho \in \left[0,1\right]$ is a $\operatorname{Beta}\left(a,b\right)$ prior, which has support in $\left[0,1\right]$. \citet{ravindran2011bayesian} mentioned that $\operatorname{Beta}\left(a,a\right)$ is a class of non-informative priors for $\rho$. However, the beta prior behaves markedly differently with different choices for $a$ and $b$, and therefore they need to be chosen carefully.

The circular distributions and priors discussed above show the diversity and utility of existing methods for modeling circular data. However, many of the priors currently in use are either heuristic or tailored to specific cases, lacking a unified framework. This motivates the need for a cohesive approach for default prior selection which can account for the unique properties of circular distributions, as presented in the next section.

\section{Methodology} \label{sec:a_principled_framework_for_prior_selection}

This section introduces a framework for constructing contraction hyperpriors for circular distributions that favors simpler circular models. The framework is presented in Section~\ref{sec:a_principled_framework_for_prior_selection.model_complexity_check} together the corresponding prior formulation framework that satisfies them. Specific priors for widely used circular distributions are proposed in Section~\ref{sec:pc_priors_for_popular_circular_distributions}.

\subsection{Penalizing complexity prior for circular models} \label{sec:a_principled_framework_for_prior_selection.model_complexity_check}

When there is insufficient information supporting the need for a complicated model, the principle of \textit{Occam's razor} \citep{mackay2003information} suggests favoring simpler alternatives. In this context, model complexity should be carefully considered when specifying prior distributions. One effective way to penalize Bayesian model complexity is through the choice of an appropriate prior, ensuring that it exerts a reasonable influence on the effective complexity of the model. Motivated by this idea, \citet{simpson2017penalising} proposed the PC prior framework, which is built upon four principles: (1) \textit{Occam's razor}; (2) \textit{measure of complexity}; (3) \textit{constant rate penalization}; and (4) \textit{user-defined scaling}.

Apart from the first principle, the \textit{measure of complexity} principle states that the prior should be constructed using a model complexity measure defined by the "distance" $d$ between the model of interest and a simpler base model. The \textit{constant rate penalization} principle specifies that the PC prior density decays at a constant rate $r \in \left(0,1\right)$ with respect to an increment $\delta$ in the distance $d$, satisfying
\begin{equation}
    \frac{p_{d}\left(d+\delta\right)}{p_{d}\left(d\right)}=r^{\delta}, \quad d,\delta \geq 0.
\end{equation}
Finally, the \textit{user-defined scaling} principle suggests that users should have some understanding of the data, or of the expected model complexity, in order to define the strength of penalization accordingly.

The PC prior framework has been applied in a range of fields. For instance, it has been used in spatial and spatio-temporal modelling \citep{fuglstad2019constructing, cabral2023controlling, rodriguez2025multivariate}, time-series analysis \citep{sorbye2017penalised}, survival modelling \citep{van2021principled}, and nonparametric regression \citep{ventrucci2016penalized}. These applications illustrate how PC priors have been used to control model complexity across different modelling contexts, motivating their extension to circular data.

In Euclidean space, statistical measures such as standard deviation and variance are intuitive and straightforward in terms of interpretability. However, as discussed in Section \ref{sec:review_on_circular_distributions_and_priors}, these quantities lose their intuitive meanings for data collected on a circular domain (or transformed into circular form) and concentration parameters are used. The parameterization of the concentration parameter varies among circular distributions but typically includes the value "0" as a special case, representing no concentration (or maximum dispersion). In most common cases, the distribution is simplified to the circular uniform distribution. It should be noted that the mean direction is not likelihood identifiable in the circular uniform distribution.

In contrast, as the concentration parameter increases, most circular distributions tend toward a point mass, representing extreme concentration around a specific direction (mean). An exception is the cardioid distribution which tends toward a cardioid curve rather than collapses into a point mass.

Although most circular distributions exhibit desirable properties when their concentration parameters approach the boundary values, the inference for these distributions can be unstable because the data are defined on a limited range ($[0, 2\pi)$). As a result, small variations in the data can correspond to large changes on the circular scale, making the concentration parameter highly sensitive. In particular, when the data deviate from the boundary cases, that is, when they are neither strongly clustered nor close to circular uniformity, the underlying model structure becomes more complex and the estimation of the concentration parameters becomes difficult and less stable.

This sensitivity and instability indicate the need for an approach to regularizing model complexity. One natural way to achieve this is to view the boundary cases as simpler circular models, which can be regarded as the “\textit{base model}” in the PC prior framework, that is, the simplest models representing minimal structure and the most straightforward interpretation. The notion of model complexity can then be described by the \textit{measure of complexity} principle, where the model complexity is quantified as the distance from the base model. We let $p\left(x\mid \xi\right)$ be the model of interest, where $x$ is the data and $\xi$ is the parameter of interest, and let $p\left(x\mid \xi_{0}\right)$ represent the base model, with $\xi = \xi_{0}$ be the corresponding parameter value at the base model. \citet{simpson2017penalising} proposed a reasonable distance measure, and in circular scenarios it can be written as
\begin{equation}
    \label{eq:kld}
    \begin{aligned}
        d\left(\xi\right) &= \sqrt{\text{KLD}\left( p\left(x\mid \xi\right) \big\Vert p\left(x\mid \xi_{0}\right) \right)} = \sqrt{\int_{0}^{2\pi} p\left(x\mid \xi\right) \log\left( \frac{p\left(x\mid \xi\right)}{p\left(x\mid \xi_{0}\right)} \right) dx},
    \end{aligned}
\end{equation}
where $\text{KLD}\left( \cdot \big\Vert \cdot \right)$ denotes the Kullback-Leibler divergence (KLD) \citep{kullback1951information} between two distributions, which is one of the most widely used quantitative measures of the difference between probability distributions.

Following the formulation of \citet{simpson2017penalising}, the Penalized Complexity (PC) prior for a model parameter $\xi$ is defined as an exponential probability density function with respect to the distance, as follows:
\begin{equation}
    \label{eq:PC_general_in_distance}
    \begin{aligned}
        p\left(d\right) = \lambda \exp\left\{ -\lambda d \right\} \Longleftrightarrow p\left(\xi\right) = \lambda \exp\left\{ -\lambda d\left(\xi\right) \right\} \left\lvert \frac{\partial d\left(\xi\right)}{\partial \xi}\right\rvert.
    \end{aligned}
\end{equation}
where $\lambda$ is the scaling parameter of the PC prior, defined such that $P_{\mathcal{PC}}\left( Q\left(\xi\right) > U \right) = \alpha$, with $Q\left(\xi\right)$ being a function of $\xi$, $U$ being a reasonable value of $Q\left(\xi\right)$ and $\alpha \in \left(0,1\right)$ a value representing probability. The formulations of $Q\left(\xi\right)$ are discussed later in this section.

\begin{figure}[!ht]
    \centering
    \includegraphics[width=\linewidth]{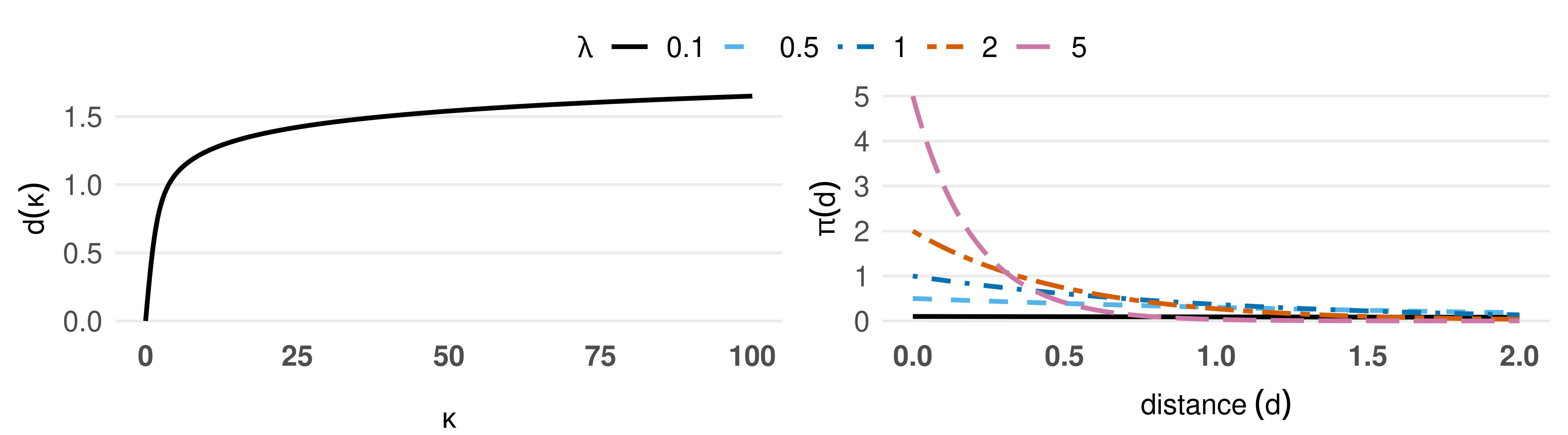}
    \caption{Distance ($d\left(\kappa\right)$) between vM distribution and circular uniform distribution (left) and the PC prior density in distance scale for different $\lambda$ values (right).}
    \label{fig:pc_density_distance_scale}
\end{figure}

To illustrate how the PC prior behaves with respect to the distance measure, \autoref{fig:pc_density_distance_scale} is presented. The left-hand-side plot shows the distance $d\left(\kappa\right)$ between the von Mises distribution and the circular uniform distribution. If we express the prior density for $\kappa$, $p\left(\kappa\right)$, in terms of distance, $p\left(d\right)$, then from the right-hand-side plot of \autoref{fig:pc_density_distance_scale}, we can observe that the prior density decreases as the distance increases. Since PC priors are exponential with respect to distance, for any concentration parameter $\xi$ of a circular distribution, the density of its PC prior will always be penalized as the distance between that circular distribution and the base model increases. Moreover, because the prior is defined to be exponential on the distance scale, this penalization occurs at a constant rate with respect to increments in distance. Therefore, the PC prior satisfies the \textit{constant rate penalization} principle, and inherently favors simpler models, penalizing the prior density exponentially as the distance $d(\xi)$ increases.
The scaling parameter, $\lambda$, of the PC prior determines the strength of penalization for deviations away from the base model. To satisfy the \textit{user-defined scaling} principle and to appropriately select the scaling parameter, users need to have some understanding of the data and its properties.
A general strategy for selecting $\lambda$ is to set $Q\left(\xi\right)=\psi$ , where $\psi \in \left[0, 1\right]$ denotes the \emph{mean resultant length}. Following \citet{ley2017modern} and \citet{lund2017package}, it is defined as
\begin{equation}
    \label{eq:mean_resultant_length}
    \begin{aligned}
        \psi &= \frac{1}{n}\sqrt{\left( \sum_{i=1}^{n}\cos\left(x_{i}\right) \right)^{2} + \left( \sum_{i=1}^{n}\sin\left(x_{i}\right) \right)^{2}},
    \end{aligned}
\end{equation}
for data $x_{i}, i=1,2,\cdots,n$. Alternative definitions of this quantity also exist, such as moving the factor $\frac{1}{n}$ inside the two squared summations \citep{mardia2009directional}. This quantity is a measure of the concentration around the circle. A value of $\psi = 0$ corresponds to a circular uniform distribution where there is no preferred direction, while $\psi=1$ indicates that all observations coincide at the same angle (maximum concentration). Under this interpretation, $P_{\mathcal{PC}}\left( Q\left(\xi\right) > U \right) = \alpha$ expresses the belief that "\textit{the probability that the mean resultant length of the data exceeds $U$, is $\alpha$}". 

This interpretation is model-independent, as the user can also express $\psi$ explicitly as a function of the model parameter $\xi$, and therefore interpret $Q\left(\xi\right)$ in terms of $\xi$. For instance, in the von Mises distribution, $\psi\left(\kappa\right) = \frac{I_1(\kappa)}{I_0(\kappa)}$, whilst for the cardioid and wrapped Cauchy distributions, $\psi$ equals to the concentration parameters $\ell$ and $\rho$ respectively. Conceptually, the mean resultant length serves as a sample-based indicator of data concentration, while the concentration parameter of a circular distribution controls the shrinkage behavior of the corresponding probability density. Therefore, the mean resultant length links the circular distribution to the observed data, incorporating the information from the data into the density function.

This unified approach covers various circular models under one framework. However, it may not be intuitive for users inexperienced with circular data. Therefore, an alternative, more intuitive approach is also suggested.

For most circular distributions, where the bounds of the concentration parameter $\xi$ correspond to the circular uniform distribution and a point mass, a suitable transformation $Q\left(\xi\right)$ can often be found. For instance, one could define $Q\left(\xi\right)$ such that circular uniform density corresponds to $Q\left(\xi\right) \to 2\pi$, and the point mass corresponds to $Q\left(\xi\right) \to 0$. Under this transformation, $Q\left(\xi\right) \in \left(0, 2\pi \right)$, making $P_{\mathcal{PC}}\left( Q\left(\xi\right) > U \right) = \alpha$ represent "the probability that the data fall outside a radian $U$ around the mean direction being equal to $\alpha$". The small values of $U$ indicate a high concentration to a point mass, whilst large values of $U$ mean that the user believes the data spread widely. This transformation is intuitive and allows users to gain insights into appropriate values for $U$ and $\alpha$ by visualizing the data.

It is worth noting that the proposed PC priors are flexible in terms of informativeness. That is, they can be either weakly or strongly informative \citep{simpson2017penalising}, depending on the user’s belief about the data. As an illustration, consider selecting the value of $\lambda$ using the mean resultant length ($Q(\xi) = \psi$). If the user believes that “the mean resultant length of the data is nearly impossible to be higher than 0.8,” then one can set $U = 0.8$ and $\alpha = 0.001$ to obtain a strongly informative prior. Conversely, if the user has no prior knowledge about the concentration of the data, $\lambda$ can be chosen by setting $U = 0.5$ and $\alpha = 0.5$, indicating uncertainty about whether the mean resultant length of the data should be larger or smaller than the median, which consequently yields a weakly informative prior. Therefore, the PC prior can always be employed regardless of whether sufficient prior information is available.

\subsection{Specific cases} \label{sec:pc_priors_for_popular_circular_distributions}

\subsubsection{Von Mises Distribution} \label{sec:priors_for_popular_circular_distribution.pc_prior_for_von_mises_distribution}
\begin{figure}[!ht]
    \centering
    \includegraphics[width=\linewidth]{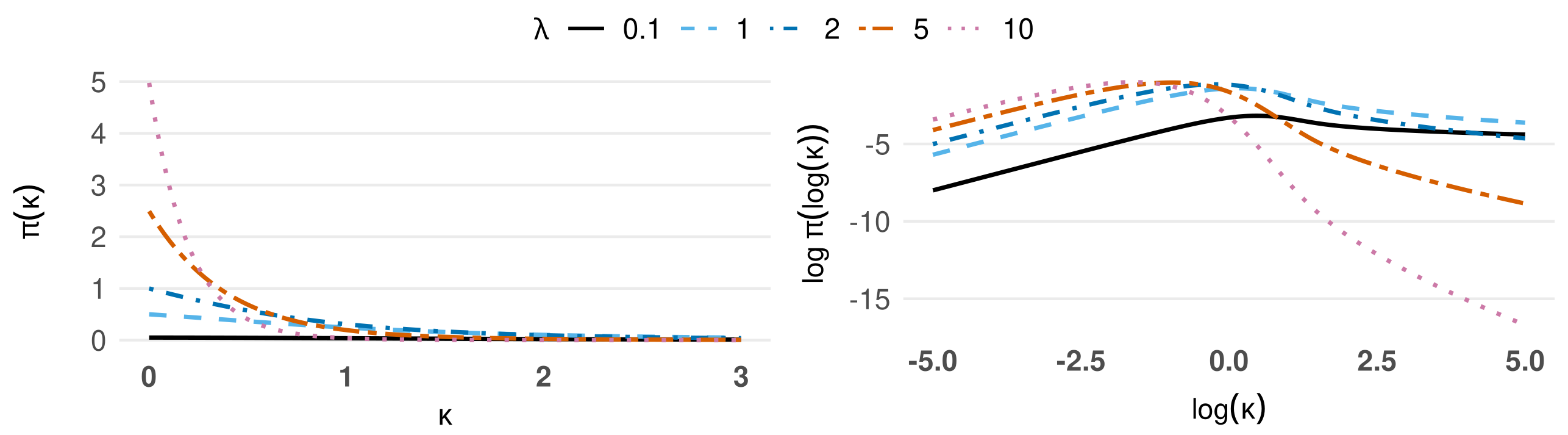}
    \caption{$\kappa_{0}=0$ base model PC prior density (left) and log-log density (right).}
    \label{fig:vm_pc_uni_pdf_and_logpdf}
\end{figure}

\begin{figure}[!ht]
    \centering
    \includegraphics[width=\linewidth]{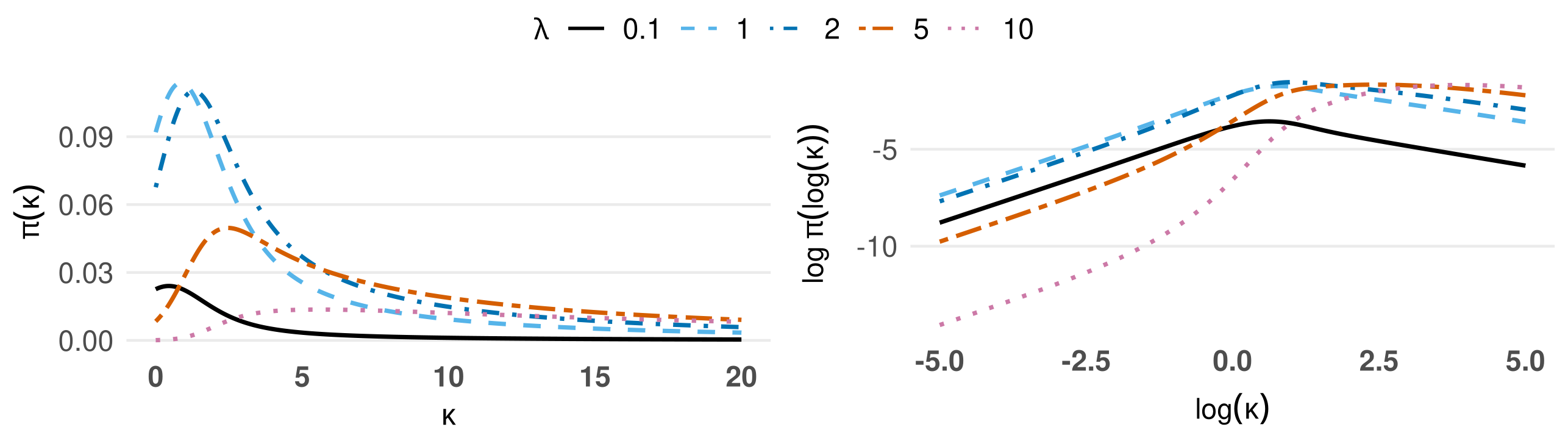}
    \caption{$\kappa_{0}\rightarrow\infty$ base model PC prior density (left) and log-log density (right).}
    \label{fig:vm_pc_pm_pdf_and_logpdf}
\end{figure}

The first step in deriving the PC prior is to compute the KLD. For the vM distribution as defined in \autoref{eq:von_Mises_density}, the KLD is given below:
\begin{equation}
    \label{eq:KLD_vM}
    \begin{aligned}
        \text{KLD}\left( \mathcal{VM} \big\Vert \mathcal{VM}_{0} \right) &= \int_{0}^{2\pi} p\left(x \mid \mu, \kappa\right) \log \left( \frac{p\left(x \mid \mu, \kappa\right)}{p\left(x \mid \mu, \kappa_{0}\right)} \right) dx \\
        &= \log \left( \mathcal{I}_{0} \left(\kappa_{0}\right) \right) - \log \left( \mathcal{I}_0 \left(\kappa\right) \right) + \frac{\kappa \mathcal{I}_{1}(\kappa)}{\mathcal{I}_{0}(\kappa)} - \frac{\kappa_{0} \mathcal{I}_{1}(\kappa)}{\mathcal{I}_{0}(\kappa)}.
    \end{aligned}
\end{equation}
Recall the behavior of the vM density with respect to changes in the concentration parameter $\kappa$ (\Cref{sec:review_on_circular_distributions_and_priors.von_mises_distribution}). Two natural choices for base model for the concentration parameter are the circular uniform distribution ($\kappa_{0}=0$), representing no concentration, and the point mass ($\kappa_{0} \gg \kappa$), representing high concentration. With these two chosen base models, the corresponding PC priors for $\kappa$ can be computed and are given in \Cref{prop:PC_kappa_uni} and \Cref{prop:PC_kappa_pm} respectively. The proofs are given in Appendix A.1.
\begin{proposition}
    \label{prop:PC_kappa_uni}
    The PC prior for concentration parameter $\kappa$ of the von Mises distribution with the base model at $\kappa_{0}=0$ has density
    \begin{equation}
        \label{eq:PC_kappa_uni_density}
        \begin{aligned}
            p(\kappa) &= \lambda \exp\left\{ - \lambda \sqrt{\frac{\kappa \mathcal{I}_{1}(\kappa)}{\mathcal{I}_{0}(\kappa)} - \log \left( \mathcal{I}_{0} \left(\kappa\right) \right)}\right\} \frac{\frac{\kappa \left( \mathcal{I}_{0} \left(\kappa\right) + \mathcal{I}_{2} \left(\kappa\right) \right)}{2\mathcal{I}_{0} \left(\kappa\right)} - \frac{\kappa \mathcal{I}_{1} \left(\kappa\right)^2}{\mathcal{I}_{0} \left(\kappa\right)^2} }{2\sqrt{\frac{\kappa \mathcal{I}_{1}(\kappa)}{\mathcal{I}_{0}(\kappa)} - \log \left( \mathcal{I}_{0} \left(\kappa\right) \right)}},
        \end{aligned}
    \end{equation}
    where $\lambda > 0$, and the corresponding CDF is
    \begin{equation}
        \label{eq:PC_kappa_uniform_cdf}
        \begin{aligned}
            F\left(\kappa\right) &= 1 - \exp\left\{ - \lambda \sqrt{\frac{\kappa \mathcal{I}_{1}(\kappa)}{\mathcal{I}_{0}(\kappa)} - \log \left( \mathcal{I}_{0} \left(\kappa\right) \right)}\right\}.
        \end{aligned}
    \end{equation}
\end{proposition}
\begin{proposition}
    \label{prop:PC_kappa_pm}
    The PC prior for concentration parameter $\kappa$ of the von Mises distribution with the base model at $\kappa_{0} \gg \kappa$ has density
    \begin{equation}
        \label{eq:PC_kappa_pm_density}
        \begin{aligned}
            p\left(\kappa\right) &= \lambda \exp\left\{ - \lambda \sqrt{1 - \frac{\mathcal{I}_{1}(\kappa)}{\mathcal{I}_{0}(\kappa)}} \right\} \frac{ \frac{\mathcal{I}_{0}(\kappa) + \mathcal{I}_{2}(\kappa)}{2\mathcal{I}_{0}(\kappa)} - \frac{\mathcal{I}_{1}(\kappa)^{2}}{\mathcal{I}_{0}(\kappa)^{2}} }{2\sqrt{1 - \frac{\mathcal{I}_{1}(\kappa)}{\mathcal{I}_{0}(\kappa)}}},
        \end{aligned}
    \end{equation}
    where $\lambda > 0$, and the corresponding CDF is
    \begin{equation}
        \label{eq:PC_kappa_pm_cdf}
        \begin{aligned}
            F\left(\kappa\right) &= \exp\left\{ - \lambda \sqrt{1 - \frac{\mathcal{I}_{1}(\kappa)}{\mathcal{I}_{0}(\kappa)}} \right\}.
        \end{aligned}
    \end{equation}
\end{proposition}
The plots for the density of both PC prior and the corresponding log-log density are presented in \autoref{fig:vm_pc_uni_pdf_and_logpdf} (prior with $\kappa_{0}=0$ base model) and \autoref{fig:vm_pc_pm_pdf_and_logpdf} (prior with $\kappa_{0}\rightarrow\infty$ base model). The plots show that both PC priors vary with the value of scaling parameter $\lambda$. For the PC prior with the base model being circular uniform density ($\kappa_{0}=0$), the prior density reaches the peak at $\kappa=0$, which indicates that this prior prefers a more uniformly distributed model. As for the PC prior with the base model being a point mass (\autoref{fig:vm_pc_pm_pdf_and_logpdf}), it assigns more density away from $\kappa=0$ when $\lambda$ increases. In other words, this PC prior believes that the data are more concentrated compared with the PC prior with circular uniform base model.
Since $\kappa \in \left[0,\infty\right)$, following the discussion on interpretable transformation in \Cref{sec:a_principled_framework_for_prior_selection.model_complexity_check}, we suggest the transformation $Q\left(\kappa\right) = \frac{2\pi}{1 + \kappa} \in \left[0, 2\pi\right]$, which represents the radian of a circle. Therefore, the scaling parameter $\lambda$ for the PC prior for the concentration parameter of von Mises distribution can be computed and calibrated by
    \begin{align}
        P\left( Q\left(\kappa\right) > U \right) &= P\left( \frac{2\pi}{1 + \kappa} > U \right) = P\left( \kappa < \frac{2\pi}{U} - 1 \right) = F\left( \frac{2\pi}{U} - 1 \right) = \alpha.
    \end{align}
Therefore, the expressions for computing $\lambda$ for PC priors for $\kappa$ are given by
\begin{equation}
    \label{eq:vm_scaling_parameter_expression}
    \begin{aligned}
        \lambda = 
        \begin{cases}
            -\frac{\log\left(1-\alpha\right)}{\sqrt{\frac{ \left(\frac{2\pi}{U} - 1\right) \mathcal{I}_{1}( \frac{2\pi}{U} - 1)}{\mathcal{I}_{0}( \frac{2\pi}{U} - 1)} - \log \left( \mathcal{I}_{0} \left( \frac{2\pi}{U} - 1\right) \right)}} \quad &\text{for base model at} \quad \kappa_{0} = 0; \\
            -\frac{\log\left(1-\alpha\right)}{\sqrt{1 - \frac{\mathcal{I}_{1}(\frac{2\pi}{U} - 1)}{\mathcal{I}_{0}(\frac{2\pi}{U} - 1)}}} \quad &\text{for base model at} \quad \kappa_{0} \rightarrow \infty.
        \end{cases}
    \end{aligned}
\end{equation}

\subsubsection{Cardioid Distribution} \label{sec:priors_for_popular_circular_distribution.pc_prior_for_cardioid_distribution}

\begin{figure}[!ht]
    \centering
    \includegraphics[width=\linewidth]{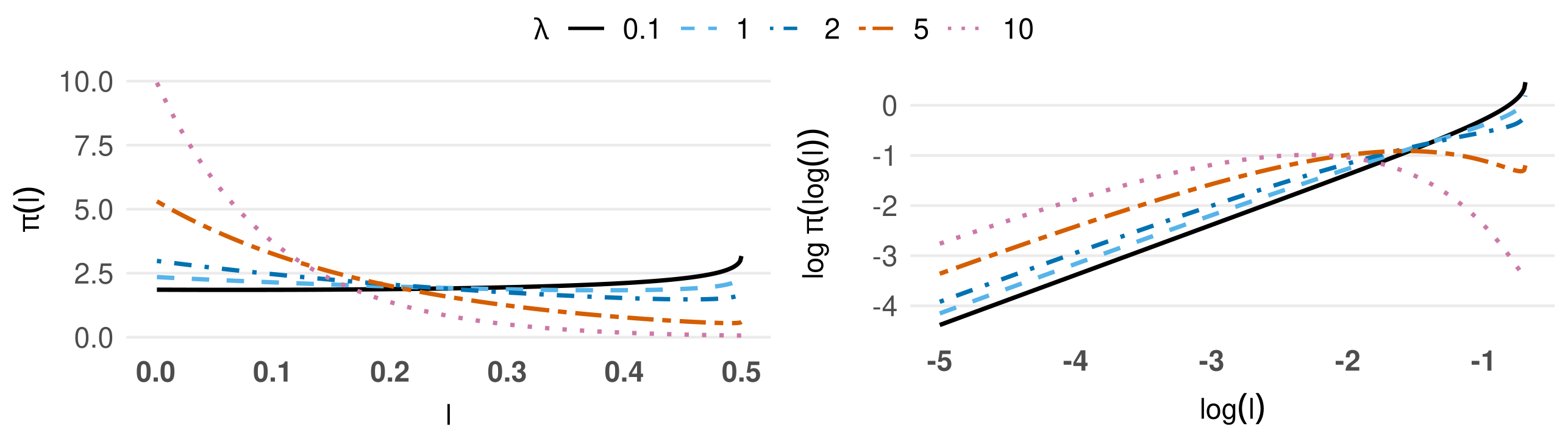}
    \caption{$\ell_{0}=0$ base model PC prior density (left) and log-log density (right).}
    \label{fig:card_pc_uni_pdf_and_logpdf}
\end{figure}

\begin{figure}[!ht]
    \centering
    \includegraphics[width=\linewidth]{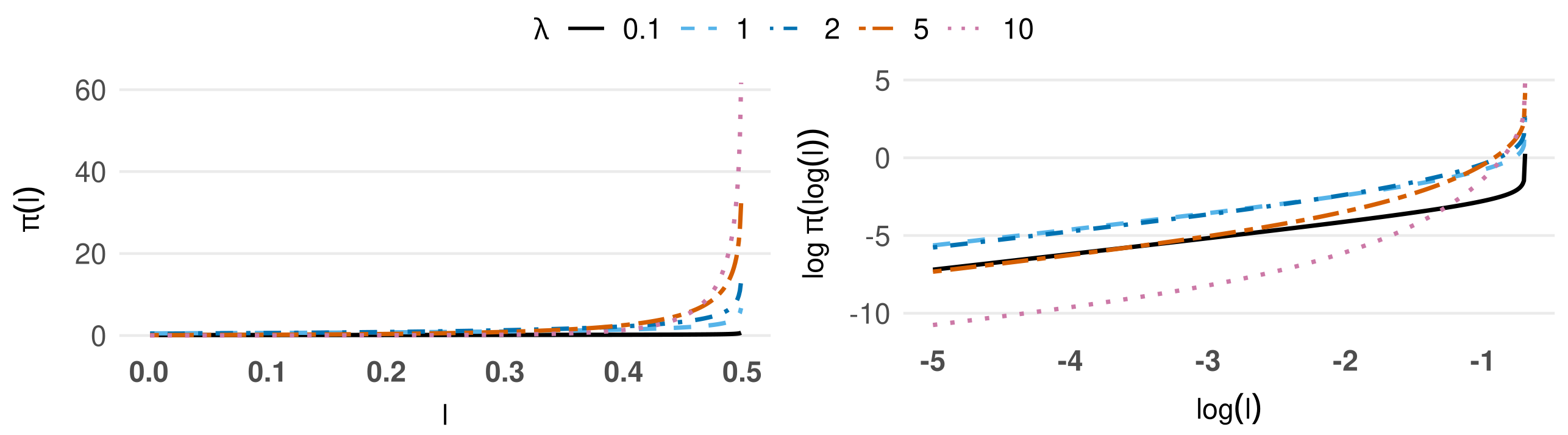}
    \caption{$\ell_{0}\rightarrow 0.5$ base model PC prior density (left) and log-log density (right).}
    \label{fig:card_pc_card_pdf_and_logpdf}
\end{figure}

For the cardioid distribution as defined in \eqref{eq:cardioid_density}, we have the KLD as
\begin{equation}
    \label{eq:KLD_cardioid}
    \begin{aligned}
        \text{KLD}\left( \mathcal{C} \big\Vert \mathcal{C}_{0} \right) &= \int_{0}^{2\pi} p\left(x \mid \mu, \ell\right) \log \left( \frac{p\left(x \mid \mu, \ell\right)}{p\left(x \mid \mu, \ell_{0}\right)} \right) dx \\
        &= 1 - \sqrt{1-4\ell^{2}} - \frac{\ell}{\ell_{0}}\left(1-\sqrt{1-4\ell_{0}^{2}}\right) + \log\left(\ell\right) - \log\left(\ell_{0}\right) \\
        &+ \frac{1}{2}\log\left( 1 + \sqrt{1-4\ell^{2}} \right) - \frac{1}{2}\log\left( 1 - \sqrt{1-4\ell^{2}} \right) \\
        &+ \frac{1}{2}\log\left( 1 - \sqrt{1-4\ell_{0}^{2}} \right) - \frac{1}{2}\log\left( 1 + \sqrt{1-4\ell_{0}^{2}} \right).
    \end{aligned}
\end{equation}
The base models for the concentration parameter $\ell$ of cardioid distribution are unusual, as $\ell$ controls the extent of the cardioid curve deviating away from the circular uniform distribution. Therefore, the two base models for $\ell$ are the circular uniform distribution $\ell_{0}=0$ and the most un-uniform cardioid curve $\left(\ell_{0} \rightarrow \frac{1}{2}\right)$, and the corresponding PC prior for $\ell$ are given in \Cref{prop:PC_ell_uni} and \Cref{prop:PC_ell_card}. The proofs are given in Appendix A.2.
\begin{proposition}
    \label{prop:PC_ell_uni}
    The PC prior for concentration parameter $\ell$ of the cardioid distribution with the base model at $\ell_{0} = 0$ has density
    \begin{equation}
        \label{eq:PC_ell_uni_density}
        \begin{aligned}
             p\left(\ell\right) &= \lambda \exp\left\{ - \lambda d\left(\ell\right)\right\} \frac{2\ell}{\left( 1 - \exp\left\{ -\lambda\sqrt{1-\log\left(2\right)} \right\} \right)\left(1+\sqrt{1-4\ell^{2}}\right)d\left(\ell\right)},
        \end{aligned}
    \end{equation}
    where $\lambda > 0$,  and 
    \begin{equation}
        \label{eq:PC_ell_uni_distance}
        \begin{aligned}
            d\left(\ell\right) &= \sqrt{1 - \sqrt{1-4\ell^{2}} + \log\left(\ell\right) + \frac{1}{2}\log\left( \frac{1 + \sqrt{1-4\ell^{2}}}{1 - \sqrt{1-4\ell^{2}}} \right)}.
        \end{aligned}
    \end{equation}
    The corresponding CDF is
    \begin{equation}
        \label{eq:PC_ell_uni_cdf}
        \begin{aligned}
            F\left(\ell\right) &= \frac{1 - \exp\left\{ - \lambda d\left(\ell\right)\right\}}{1 - \exp\left\{ -\lambda  \sqrt{1 - \log\left(2\right)} \right\}}.
        \end{aligned}
    \end{equation}
\end{proposition}
\begin{proposition}
    \label{prop:PC_ell_card}
    The PC prior for concentration parameter $\ell$ of the cardioid distribution with the base model at $\ell_{0} \rightarrow \frac{1}{2}$ has density
    \begin{equation}
        \label{eq:PC_ell_card_density}
        \begin{aligned}
             p\left(\ell\right) &= \lambda \exp\left\{ - \lambda d\left(\ell\right)\right\} \frac{2\ell + \sqrt{1-4\ell^{2}} - 1}{2\ell d\left(\ell\right)},
        \end{aligned}
    \end{equation}
    where $\lambda > 0$,  and 
    \begin{equation}
        \label{eq:PC_ell_card_distance}
        \begin{aligned}
            d\left(\ell\right) &= \sqrt{1 + \log\left(2\right) - 2\ell - \sqrt{1-4\ell^{2}} + \log\left(\ell\right) + \frac{1}{2}\log\left( \frac{1 + \sqrt{1-4\ell^{2}}}{1 - \sqrt{1-4\ell^{2}}} \right)}.
        \end{aligned}
    \end{equation}
    The corresponding CDF is
    \begin{equation}
        \label{eq:PC_ell_card_cdf}
        \begin{aligned}
            F\left(\ell\right) &= \exp\left\{ - \lambda d\left(\ell\right)\right\}.
        \end{aligned}
    \end{equation}
\end{proposition}
The plots for the density of both PC prior and the respected log-log density are presented in \autoref{fig:card_pc_uni_pdf_and_logpdf} (prior with $\ell_{0}=0$ base model) and \autoref{fig:card_pc_card_pdf_and_logpdf} (prior with $\ell_{0}\rightarrow 0.5$ base model). \autoref{fig:card_pc_uni_pdf_and_logpdf} illustrates that the PC prior with a uniform base model assigns higher density at $\ell = 0$ for larger values of $\lambda$. However, there is always a sharp increase in density as $\ell \to 0.5^{-}$. In contrast, for the PC prior with $\ell_{0} \to 0.5$ as the base model (\autoref{fig:card_pc_card_pdf_and_logpdf}), the prior density increases monotonically with $\ell$. Additionally, the rate of increase becomes steeper as $\ell$ approaches 0.5.
The parameter $\ell$ serves as a shape-concentration parameter, controlling the "cardiodity" of the density. Unlike the von Mises (vM) and wrapped Cauchy (WC) distributions, it is not appropriate to define $Q\left(\ell\right)$ using similar ideas due to the unique role of $\ell$ in determining the cardioid shape. Instead, we propose the transformation $Q\left(\ell\right) = 2\ell \in \left(0, 1\right)$, which provides a meaningful interpretation as the "cardiodity rate" from 0 to 1. Using this transformation, the scaling parameter $\lambda$ in \autoref{eq:PC_ell_card_density} can be calculated by:
\begin{equation}
    \label{eq:cardioid_compute_scaling_parameter}
    \begin{aligned}
        P\left( Q\left(\ell\right) > U \right) &= P\left( 2\ell > U \right) = P\left( \ell > \frac{U}{2} \right) = 1 - F\left( \frac{U}{2} \right) = \alpha
    \end{aligned}
\end{equation}
This transformation ensures an intuitive and interpretable way to set $U$, representing the threshold for the "cardiodity rate" and $\alpha$, which controls the weight assigned to the tail of the PC prior.

Through calculation, the expressions for computing $\lambda$ of PC priors of $\kappa$ are given by
\begin{equation}
    \label{eq:card_scaling_parameter_expression}
    \begin{aligned}
        \lambda = 
        \begin{cases}
            -\frac{\log\left( \alpha + \left(1-\alpha\right)\exp\left\{ -\lambda \sqrt{1-\log 2} \right\} \right)}{d\left( \frac{U}{2} \right)} \quad &\text{for base model at} \quad \ell_{0} = 0; \\
            -\frac{\log\left( 1 - \alpha \right)}{d\left( \frac{U}{2} \right)} \quad &\text{for base model at} \quad \ell_{0} \rightarrow \frac{1}{2},
        \end{cases}
    \end{aligned}
\end{equation}
where $d\left( \cdot \right)$ for both cases are given in \Cref{prop:PC_ell_uni} and \Cref{prop:PC_ell_card} respectively.

\subsubsection{Wrapped Cauchy Distribution} \label{sec:priors_for_popular_circular_distribution.pc_prior_for_wrapped_cauchy_distribution}

\begin{figure}[!ht]
  \centering
    \includegraphics[width=\textwidth]{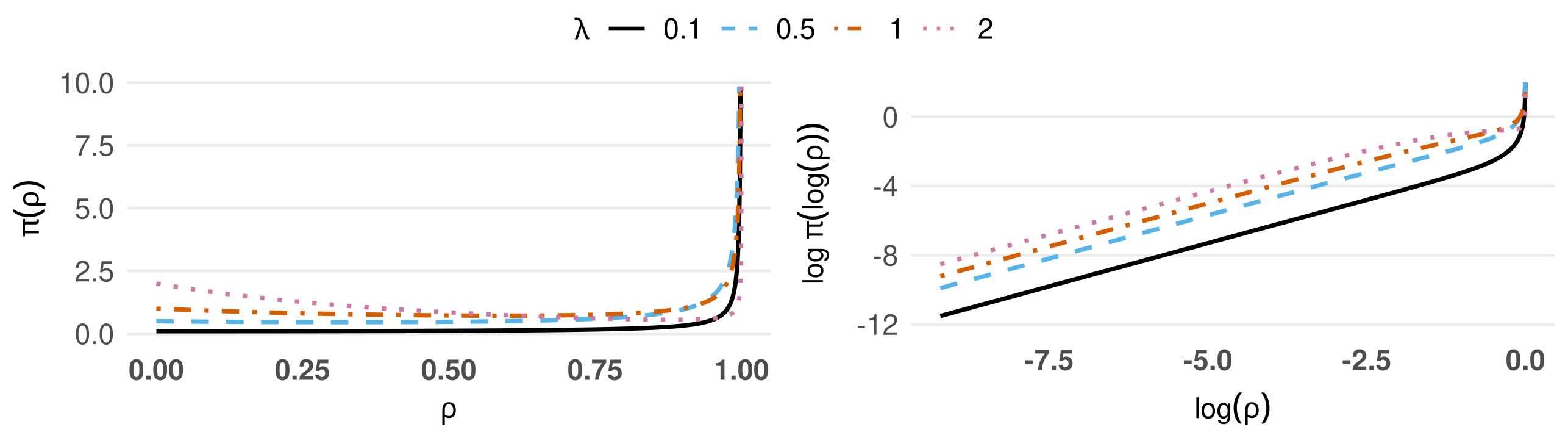}
    \caption{PC prior (for $\rho$) density (left) and log-log density (right).}
    \label{fig:wc_pc_pdf_and_logpdf}
\end{figure}

Recall that $\rho=1$ is not well defined for the wrapped Cauchy density (\Cref{sec:review_on_circular_distributions_and_priors.wrapped_cauchy_distribution}), and also that the WC distribution is preferred as a model with less concentration, when compared with vM distribution. WC distribution is thus usually employed when we believe the data are not strongly concentrated. Thus, it is more reasonable to consider the circular uniform distribution ($\rho_{0}=0$) as the only base model for concentration parameter $\rho$ of the WC distribution. Therefore, the KLD between $p \left( x \mid \mu, \rho \right)$ and $p \left( x \mid \mu, \rho_{0}=0 \right)$ is 
\begin{equation}
    \label{eq:KLD_WC_uniform_derivation}
    \begin{aligned}
        \text{KLD}\left( \mathcal{WC} \big\Vert \mathcal{WC}_{0} \right) &= \int p \left( x \mid \mu, \rho \right) \log \left( \frac{p \left( x \mid \mu, \rho \right)}{p \left( x \mid \mu, \rho_{0} \right)} \right) dx \\
        &= - \log\left(1-\rho^2\right) - \log \left( 1 - \rho_{0}^2 \right) + 2\log \left(1-\rho_{0}\rho\right) \\
        &= - \log\left(1-\rho^2\right)
    \end{aligned}
\end{equation}
With this result, the PC prior for $\rho$ is given in \Cref{prop:PC_rho}.
\begin{proposition}
    \label{prop:PC_rho}
    The PC prior for concentration parameter $\rho$ of the wrapped Cauchy distribution with the base model at $\rho_{0}=0$ has density
    \begin{equation}
        \label{eq:PC_rho_density}
        \begin{aligned}
            p\left(\rho\right) &= \lambda \exp\left\{ - \lambda \sqrt{- \log \left( 1 - \rho^{2} \right)}\right\} \frac{\rho}{\left(1-\rho^2\right)\sqrt{- \log \left( 1 - \rho^{2} \right)}},
        \end{aligned}
    \end{equation}
    where $\lambda > 0$, and the corresponding CDF is
    \begin{equation}
        \label{eq:PC_rho_cdf}
        \begin{aligned}
            F\left(\rho\right) &= 1- \exp\left\{ - \lambda \sqrt{- \log \left( 1 - \rho^{2} \right)}\right\}.
        \end{aligned}
    \end{equation}
\end{proposition}
The proof is provided in Appendix A.3, and the density and log-log density plots of this PC prior are shown in \autoref{fig:wc_pc_pdf_and_logpdf}. The density plot indicates that higher values of $\lambda$ result in greater prior density at $\rho = 0$, and the prior density increases sharply as $\rho$ approaches 1. In the log-log scale, the prior behavior reveals a consistent trend across different values of $\lambda$, suggesting similar characteristics despite variations in $\lambda$.
Although the point mass boundary for WC distribution $\rho=1$ is not well-defined, $\rho$ still controls the concentration behavior of the distribution from uniform to approximate point mass. Therefore, to compute the scaling parameter of the PC prior, the same idea of interpretable transformation for a parameter could be employed. For $\rho \in \left[0,1\right)$, we propose to use the transformation $Q\left(\rho\right) = 2\pi\left(1-\rho\right) \in \left[0,2\pi\right]$. Now, $Q\left(\rho\right) \rightarrow 2\pi$ when $\rho \rightarrow 0$ and $Q\left(\rho\right) \rightarrow 0$ when $\rho \rightarrow 1$. Then, by solving $P\left( Q\left(\rho\right) > U \right) = F\left( 1 - \frac{U}{2\pi} \right) = \alpha$, we have $\lambda = -\log\left(1-\alpha\right)/\sqrt{-\log\left( \frac{U}{\pi} - \frac{U^{2}}{4\pi^{2}} \right)}$.

\section{Simulation Study and Comparisons} \label{sec:comparison_and_investigation_of_common_priors}

The logic behind our proposed prior selection framework is to assess whether a prior favors simpler models and discourages unnecessarily complex specifications.

The logic behind our proposed prior selection framework is to assess whether a prior favors simpler models and discourages unnecessarily complex specifications. To further illustrate the way of using our framework, and to demonstrate the advantages of employing PC priors for circular models, this section compares the proposed priors in \Cref{sec:priors_for_popular_circular_distribution.pc_prior_for_von_mises_distribution} with existing priors from the literature. The comparison focuses on their behavior in both the parameter scale and the distance scale (\Cref{sec:comparison_and_investigation_of_common_priors.properties_of_priors}). Additionally, a comprehensive set of simulation studies is presented in \Cref{sec:comparison_and_investigation_of_common_priors.simulation_study}.

\subsection{Properties of PC and other common priors} \label{sec:comparison_and_investigation_of_common_priors.properties_of_priors}
In this section we illustrate the behavior of the proposed PC prior as well as other priors from the literature for the three cases models under consideration.
\subsubsection{Von Mises Distribution} \label{sec:comparison_and_investigation_of_common_priors.von_mises_distribution}

\begin{figure}[!ht]
  \centering
    \includegraphics[width=\textwidth]{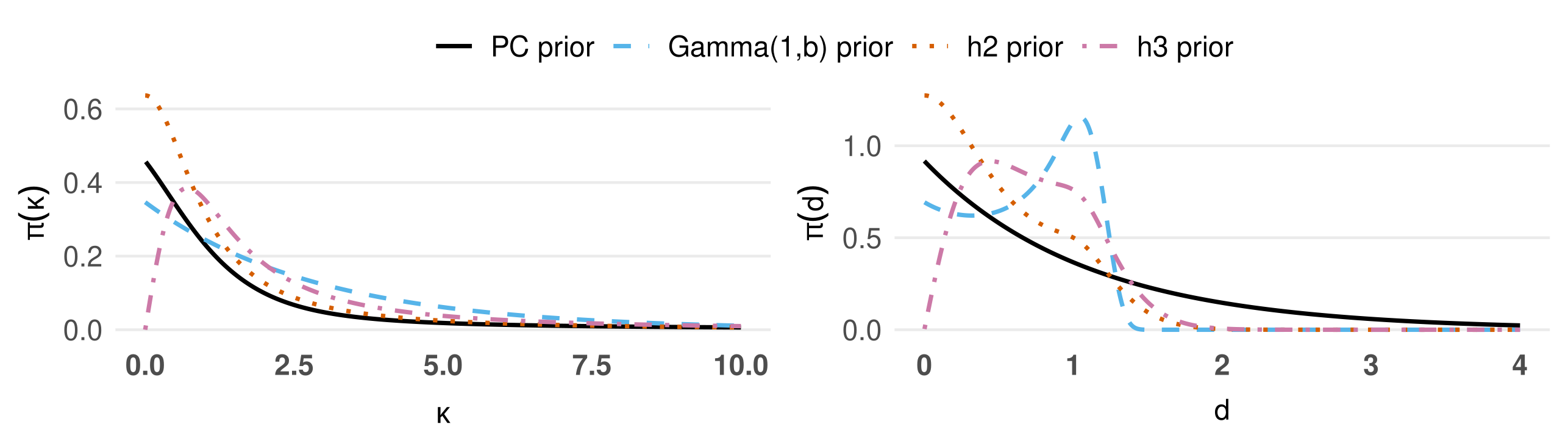}
    \caption{Priors for $\kappa$ in parameter (left) and distance scale (right) with the base model at $\kappa_{0}=0$. The solid line is the $\operatorname{PC}\left( \lambda=0.92 \right)$ prior, the dashed line is the $\operatorname{Gamma}\left(1,0.34\right)$, the dotted line is the $h_{2}$ prior and the dot-dash line is the $h_{3}$ prior.}
    \label{fig:compare_vm_uni_priors}
\end{figure}

\autoref{fig:compare_vm_uni_priors} illustrates the behaviors of the PC prior, the popular $\operatorname{Gamma}(1, b)$ prior, $h_{2}$ prior \autoref{eq:von_mises_kappa_h2_and_h3_prior}, and $h_{3}$ prior \autoref{eq:von_mises_kappa_h2_and_h3_prior} in both the parameter scale and the distance scale. The distance measure is defined between the von Mises (vM) density and the circular uniform density. For fairness, the scaling parameter $\lambda$ of the PC prior and the rate parameter $b$ of the $\operatorname{Gamma}(1, b)$ prior are calibrated so that both priors share the same median.

From the plot of the density on the distance scale, it is evident that the $h_{3}$ prior assigns zero density at $d = 0$, indicating that it never suggests the data being uniform. This behavior implies a preference for more complex models. The $\operatorname{Gamma}\left(1, b\right)$ prior has reasonable density at the base model, whereas it exhibits a non-monotonic density with a global maximum around $d \approx 1$ rather than at $d = 0$, which indicates limited contraction toward the base model. The $h_{2}$ prior does not penalize the density exponentially with respect to distance by nature; however, it has a reasonable behavior with respect to distance, that is, its density peaks at $d = 0$ and decreases monotonically as $d \to \infty$.
\begin{figure}[!ht]
  \centering
    \includegraphics[width=\textwidth]{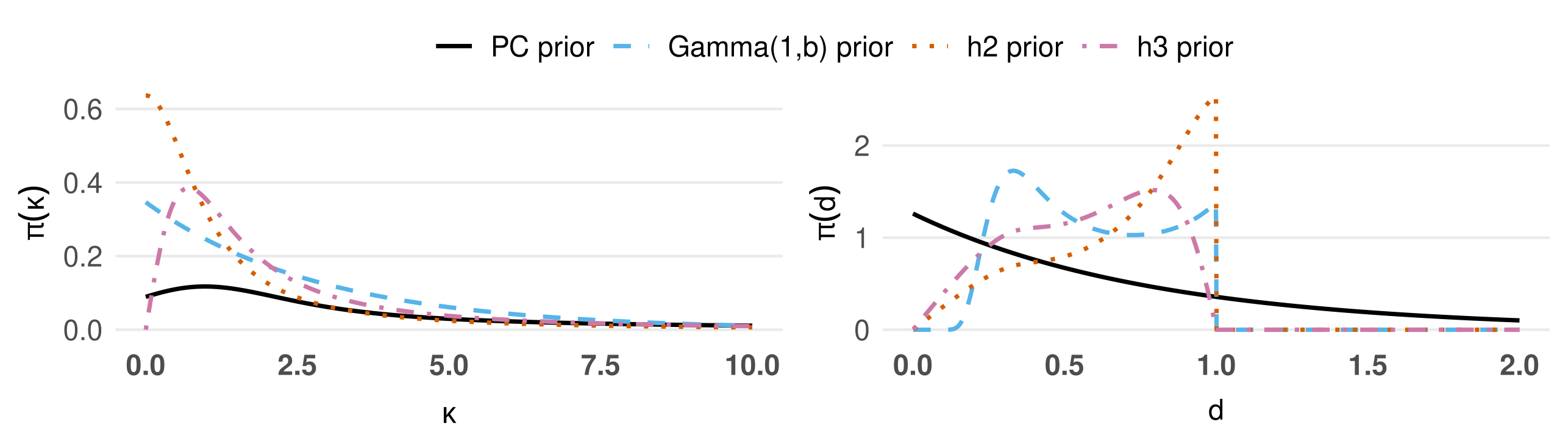}
    \caption{Priors for $\kappa$ in parameter (left) and distance scale with the base model at $\kappa_{0}\rightarrow \infty$(right). The solid line is the $\operatorname{PC}\left( \lambda=1.26 \right)$ prior, the dashed line is the $\operatorname{Gamma}\left(1,0.34\right)$, the dotted line is the $h_{2}$ prior and the dot-dash line is the $h_{3}$ prior.}
    \label{fig:compare_vm_pm_priors}
\end{figure}

When the base model is at point mass ($\kappa_{0} \to \infty$), \autoref{fig:compare_vm_pm_priors} reveals that only the PC prior has non-zero density at the simplest case ($d = 0$), whereas the $\operatorname{Gamma}\left(1, b\right)$, $h_{2}$, and $h_{3}$ priors assign zero density to $d = 0$, which limits their ability to regularize toward the base model in this scenario. Nonetheless, except for the PC prior, the other priors appear to have non-smooth density curves, with zero density for $d > 1$. In other words, these priors cannot support a model that has moderate distance from the point mass either.

Recall that the $h_{2}$ and $h_{3}$ priors are constructed based on the MML criterion, which is intended to penalize model complexity. From the plots given above, we can see that the MML-based priors assign reasonable density to small distances from the base models but do not always allow the model to reach the boundary when the distance measure is defined by \autoref{eq:kld}.

Based on these observations, we recommend using the PC prior for the concentration parameter $\kappa$ of the von Mises distribution, particularly when limited prior information or belief is available. However, in cases where the data exhibit low concentration, the $h_{2}$ prior can also be a reasonable alternative.

\subsubsection{Cardioid Distribution} \label{sec:comparison_and_investigation_of_common_priors.cardioid_distribution}

\begin{figure}[!ht]
  \centering
  \begin{minipage}[b]{0.48\textwidth}
    \includegraphics[width=\textwidth]{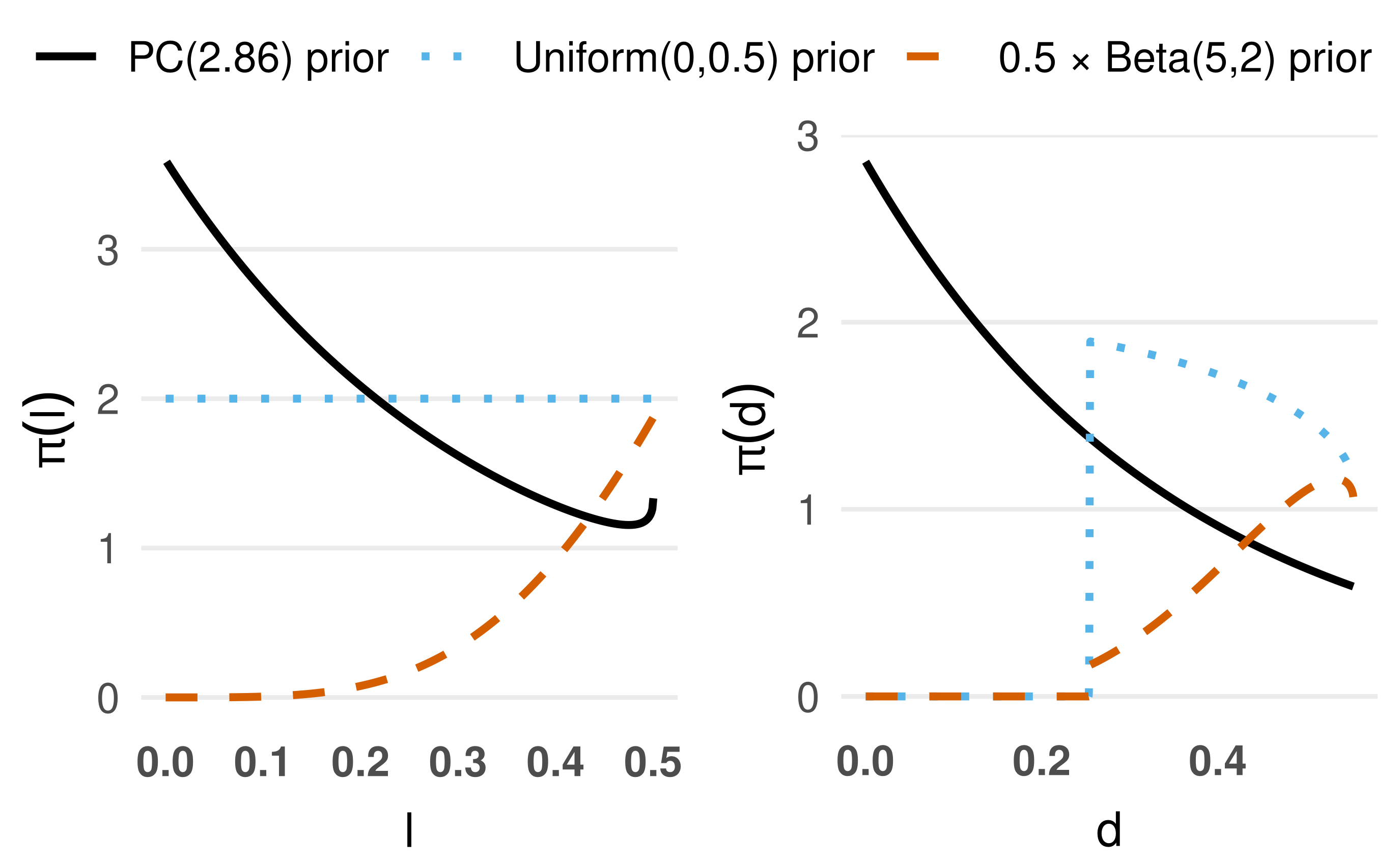}
    \caption{Priors for $\ell$ in parameter (left) and distance scale with the base model at $\ell_{0}=0$ (right). The solid line is the $\operatorname{PC}\left( \lambda=2.86 \right)$ prior, the dashed line is the $\operatorname{Beta}\left(5,2\right)$ prior and the dotted line is the uniform prior.}
    \label{fig:compare_card_uni_priors}
  \end{minipage}
  \hfill
  \begin{minipage}[b]{0.48\textwidth}
    \includegraphics[width=\textwidth]{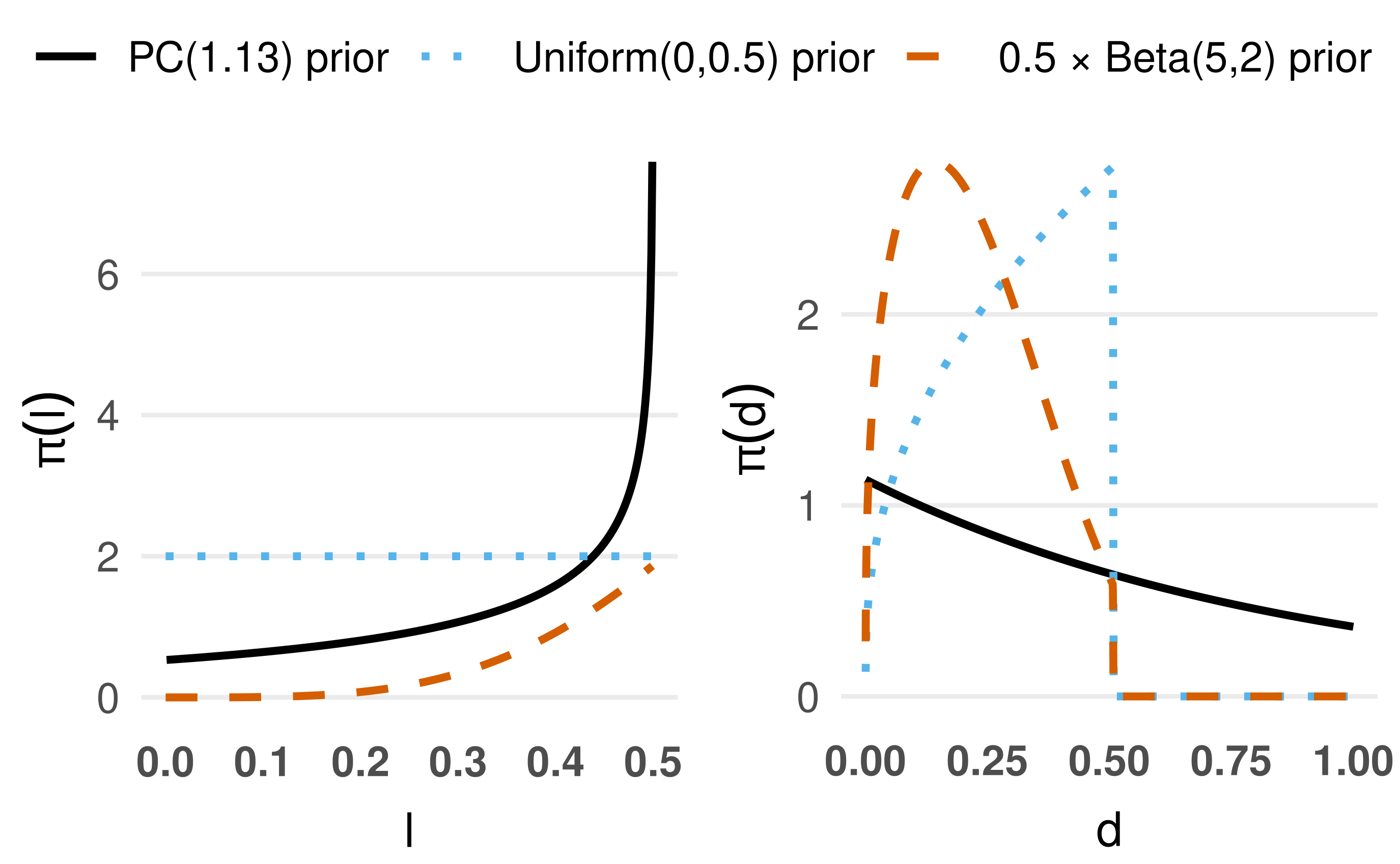}
    \caption{Priors for $\ell$ in parameter (left) and distance scale with the base model at $\ell_{0}\rightarrow 0.5$ (right). The solid line is the $\operatorname{PC}\left( \lambda=1.13 \right)$ prior, the dashed line is the $\operatorname{Beta}\left(5,2\right)$ prior and the dotted line is the uniform prior.}
    \label{fig:compare_card_card_priors}
  \end{minipage}
\end{figure}

\autoref{fig:compare_card_uni_priors} and \autoref{fig:compare_card_card_priors} show the boundary behaviors for the priors for parameter $\ell$ with the base models being at $\ell_{0}=0$ and $\ell_{0}\rightarrow 0.5$ respectively. The plots conclude that only PC priors perform well at the boundaries, assigning non-zero density at $d=0$, whilst the density of uniform and $0.5 \times \operatorname{Beta}$ priors are not even smooth. This feature illustrates the possible inherent caveat when priors are specified to be smooth and well-behaved on the parameter scale.

\subsubsection{Wrapped Cauchy Distribution} \label{sec:comparison_and_investigation_of_common_priors.wrapped_cauchy_distribution}

\begin{figure}[!ht]
  \centering
    \includegraphics[width=\textwidth]{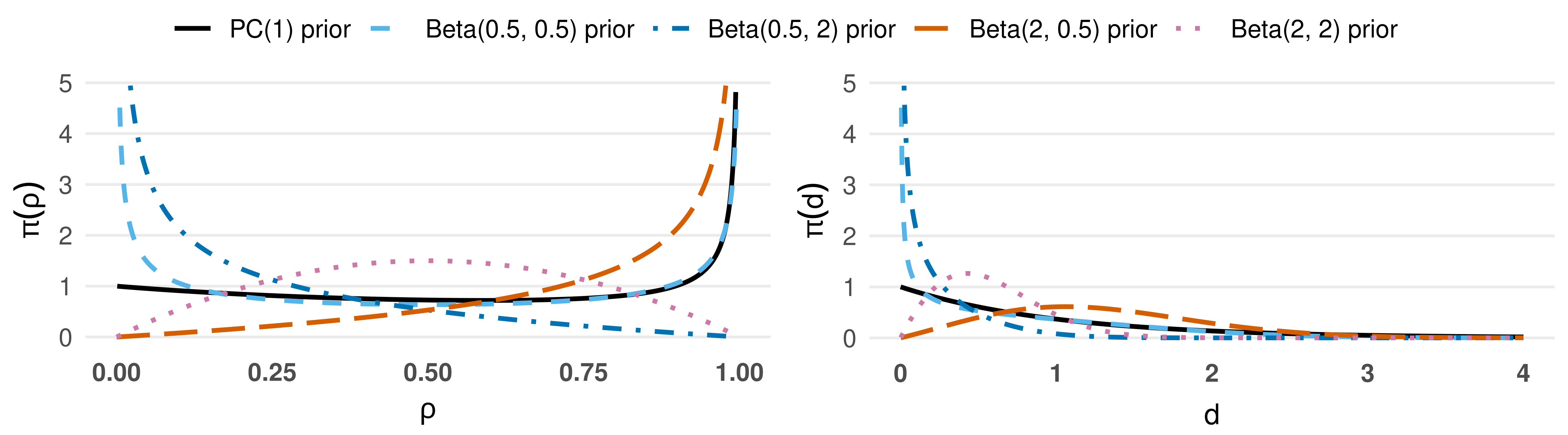}
    \caption{Beta prior and PC ($\lambda=1$) prior density in parameter ($p\left(\rho\right)$, left) and distance scale ($p\left(d\right)$, right).}
    \label{fig:compare_wc_uni_priors}
\end{figure}

The Beta prior is one of the most commonly used choices for the concentration parameter $\rho$ of the wrapped Cauchy distribution. The prior density on both the parameter scale and the distance scale is illustrated in \autoref{fig:compare_wc_uni_priors}. The behavior of the Beta prior is particularly interesting, as it exhibits different trends near the two boundaries depending on the choice of the $a$ and $b$ hyperparameters. The right-hand-side plot shows that the Beta prior assigns non-zero density at $d = 0$ only when $a < 1$, so that it does not exclude the base model a priori.
Therefore, a Beta prior with $a < 1$ can be a suitable prior choice for $\rho$. When users have sufficient information to set an informative prior, a Beta prior with $a < 1$ is a reasonable choice, as it maintains the ability to prevent "overconfidence". However, selection of the hyperparameter $b$ significantly influences the behavior of the Beta prior. Thus, we should always be careful when employing this beta prior in practice.

\subsection{Simulation Study} \label{sec:comparison_and_investigation_of_common_priors.simulation_study}

In former subsections, we illustrate the way of employing our proposed framework, and the behavior of the discussed priors are illustrated and examined. In this section, we further show the performance of the priors and the robustness of the proposed circular PC priors through comprehensive simulation studies. The simulation study is conducted through comparing the posteriors of the concentration parameter $\xi$ ($\kappa$ for von Mises distribution, $\ell$ for cardioid distribution and $\rho$ for wrapped Cauchy distribution) under different priors. For consistency, $p\left(\mu\right)$ is chosen to be the circular uniform density, thus, the joint posterior is $p\left(\mu,\xi \mid x\right) \propto p\left(x \mid \mu, \xi \right) p\left(\mu\right)p\left(\kappa\right)$.

In the study, posterior samples are obtained using the No-U-Turn Sampler (NUTS), an adaptive variant of Hamiltonian Monte Carlo (HMC), implemented through the Stan interface \citep{carpenter2017stan, stan_manual}. Sampling is performed using the rstan package version 2.32.7 together with Stan version 2.32.2 to ensure reproducibility. For consistency, the data are sampled from $p\left(x \mid \mu, \xi \right)$ (vM, cardioid and WC distribution respectively) with $N=100, 300, 1000$ sample sizes and location parameter $\mu=\pi$. In addition, the result is obtained by averaging over 100 replicates of the experiment. For reproducibility, the random seed is fixed at 520 across all experiments, with individual seeds for the 100 replicates ranging sequentially from 521 to 620. The code used in the simulation study is available at \url{https://github.com/XiangYEstats/PCpriors-circular}.

The convergence behavior of the model is assessed using the Gelman–Rubin convergence diagnostic, $\hat{R}$ \citep{gelman1992inference} (see details in the \textit{Posterior Analysis} section of \citealt{stan_manual}). An $\hat{R}$ value close to 1 indicates good model convergence. Models with $\hat{R} > 1.01$ raise convergence concerns, and $\hat{R} > 1.1$ is generally considered unacceptable.

\subsection{Von Mises Distribution}

For $p\left(\kappa\right)$ we consider the $\operatorname{Gamma}\left(1,b\right)$ (exponential) priors, the $h_{2}$ and the $h_{3}$ prior (\autoref{eq:von_mises_kappa_h2_and_h3_prior}, MML priors), and the PC priors with uniform base model (PCU, \autoref{eq:PC_kappa_uni_density}) and point mass base model (PCP, \autoref{eq:PC_kappa_pm_density}). The data for the simulation study are sampled with: $\kappa=0.02, 1, 7, 59$; The hyperparameter for $\operatorname{Gamma}\left(1,b\right)$ prior has values of $b=0.01, 0.05, 0.1, 1, 5$; The scaling parameters for both PCU and PCP priors are set with $U=\frac{\pi}{2}$, and $\alpha = 0.01, 0.1, 0.3, 0.5, 0.7, 0.9, 0.99$.
\begin{figure}[!ht]
  \centering
    \includegraphics[width=\textwidth]{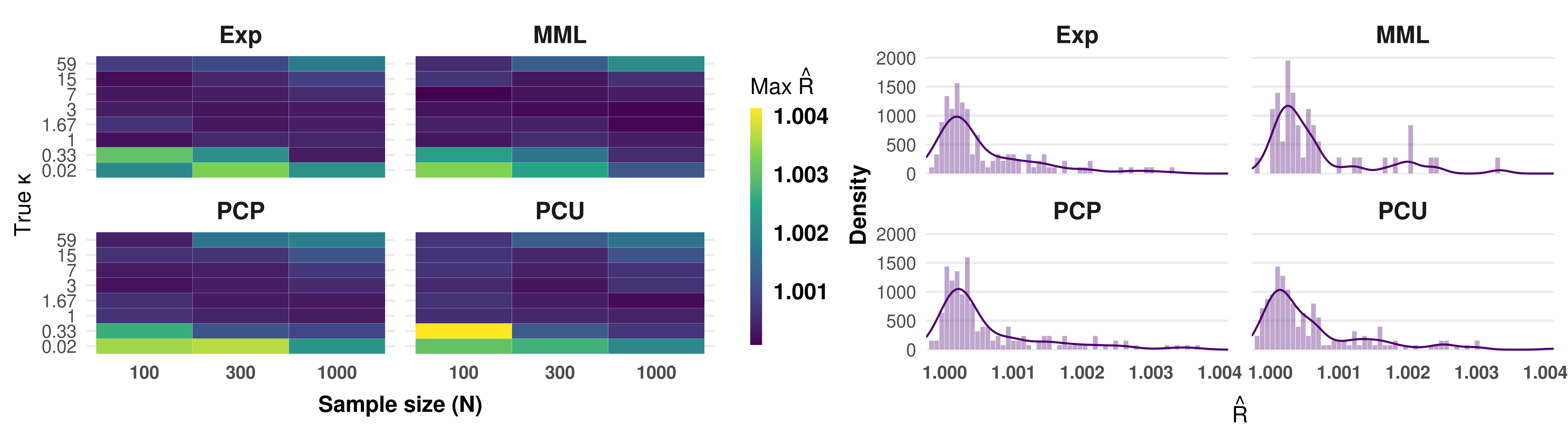}
    \caption{$\hat{R}$ values for $\kappa$ in the simulation study. The left plot shows the maximum $\hat{R}$ for each setting, and the right plot shows the $\hat{R}$ density by prior. For each setting, the $\hat{R}$ values for PCU, PCP, and exponential priors are averaged over scaling parameter values, and the $h_{2}$ and $h_{3}$ priors are averaged and presented under "MML".}
    \label{fig:rhat_vm}
\end{figure}
The $\hat{R}$ values for this simulation study are shown in \autoref{fig:rhat_vm}. The plots indicate that all priors exhibit good convergence behavior when using the Stan model, with the maximum $\hat{R}$ value being less than 1.004.

\begin{figure}[!ht]
  \centering
    \includegraphics[width=1\textwidth]{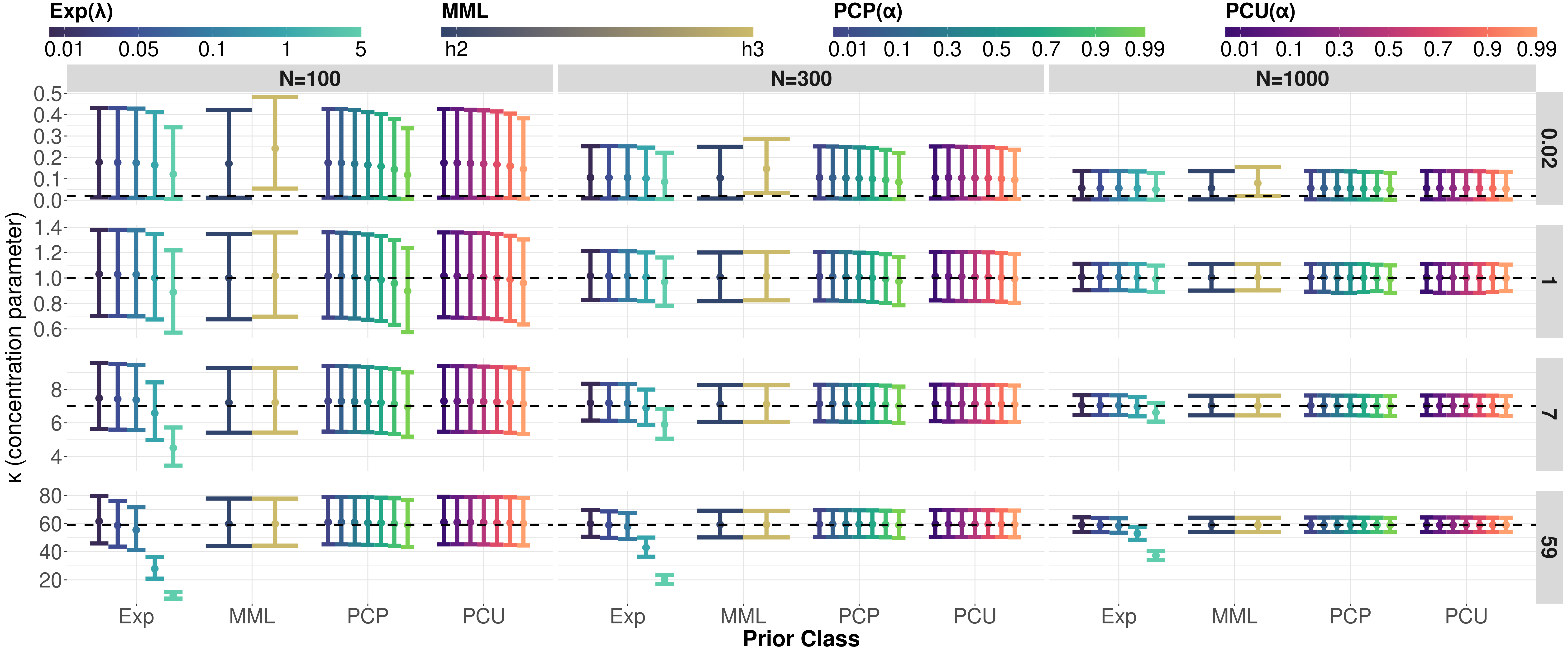}
    \caption{Simulation study for priors on the concentration parameter $\kappa$ of the von Mises distribution across different sample sizes and true $\kappa$ values. The horizontal dashed line indicates the true $\kappa$, with points showing posterior means and vertical bars indicating 95\% posterior credible intervals for each prior class.}
    \label{fig:vm_simulation}
\end{figure}

From the results in \autoref{fig:vm_simulation}, we can see that both PCU and PCP priors perform consistently across different values of $\alpha$ (resulting in different values of $\lambda$), whereas the posterior for $\operatorname{Gamma}\left(1,b\right)$ priors behaves differently, especially when the sample size is small and the concentration is high. This indicates that the $\operatorname{Gamma}\left(1,b\right)$ prior is sensitive to the choice of hyperparameter, whereas the PC priors show less sensitivity to the values of $U$ and $\alpha$. Notably, the $h_{2}$ prior also performs well in fitting the model. However, based on the discussion in \Cref{sec:comparison_and_investigation_of_common_priors.von_mises_distribution}, it is not recommended to employ the $h_{2}$ prior when the data are highly concentrated, as it may favor models that are excessively distant from the base model.

\subsection{Cardioid Distribution}
For $p\left(\ell\right)$ we consider the $\operatorname{Uniform}\left(0,0.5\right)$ prior (UNI), the $0.5\times\operatorname{Beta}\left(a,b\right)$ prior and the PC priors with the uniform base model (PCU, Equation 3.13) and the cardioid curve base model (PCC, Equation 3.16). The data are sampled with $\ell=0, 0.01, 0.3, 0.49$; The hyperparameter for $0.5\times\operatorname{Beta}\left(a,b\right)$ prior has values of $a, b \in \left\{0.5, 1, 2, 5\right\}$; The scaling parameters for both PCC and PCU priors are set with $U=0.5$, but with $\alpha = 0.01$, 0.1, 0.3, 0.5, 0.7, 0.9, 0.99 for PCC prior and $\alpha = 0.01$, 0.1, 0.2, 0.3, 0.4, 0.5 for PCU prior, since $\lambda$ takes very small values when $\alpha > 0.5$ for PCU prior.

\begin{figure}[!ht]
  \centering
    \includegraphics[width=\textwidth]{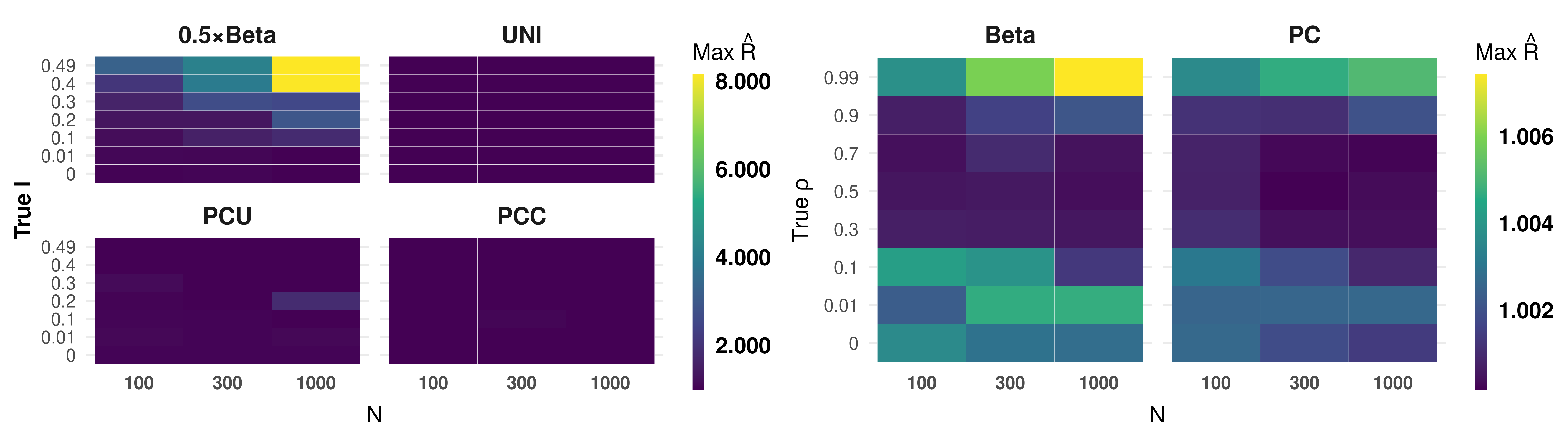}
    \caption{$\hat{R}$ values for $\ell$ (left) and $\rho$ (right) in the simulation study. The plots show the maximum $\hat{R}$ for each setting For each setting, the $\hat{R}$ values for priors with scaling parameter are averaged over scaling parameter values.}
    \label{fig:rhat_card_and_wc}
\end{figure}

The $\hat{R}$ values for this simulation study are shown in the left-hand side of \autoref{fig:rhat_card_and_wc}. The plot shows that the $0.5\times\operatorname{Beta}(a,b)$ prior does not converge in the Stan model for large $\ell$ values and large sample sizes, whereas the $\operatorname{Uniform}(0,0.5)$, PCC, and PCU priors converge successfully.
\begin{figure}[!ht]
  \centering
    \includegraphics[width=\textwidth]{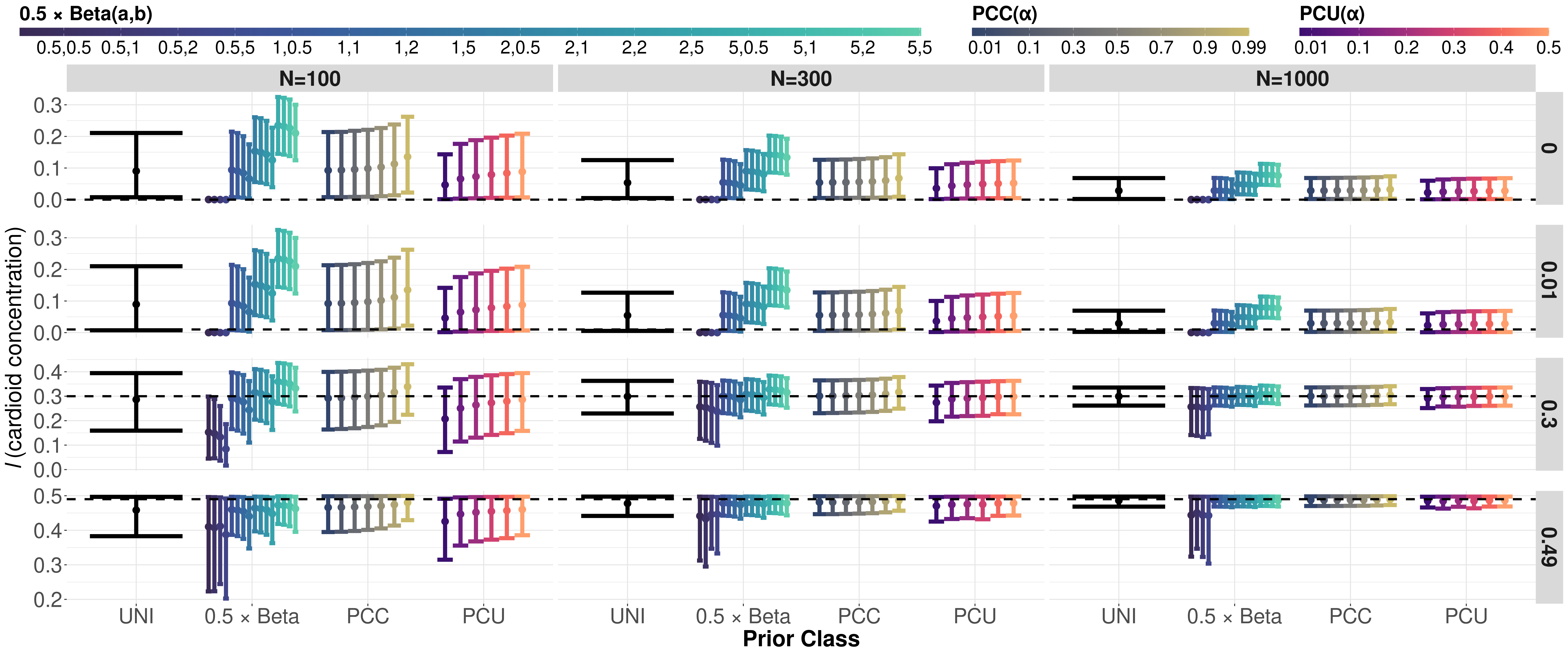}
    \caption{Simulation study for priors on the concentration parameter $\ell$ of the cardioid distribution across different sample sizes and true $\ell$ values. The horizontal dashed line indicates the true $\ell$, with points showing posterior means and vertical bars indicating 95\% posterior credible intervals for each prior class.}
    \label{fig:cardioid_simulation}
\end{figure}

From the results given in \autoref{fig:cardioid_simulation}, we could observe that both PCU and PCC priors have much more stable and accurate performance than the $0.5\times\operatorname{Beta}\left(a,b\right)$ prior. Moreover, the PCU prior performs better than the PCC prior around the $\ell=0$ boundary, whilst the PCC prior naturally perform better around the $\ell \rightarrow 0.5$ boundary. This fact demonstrates the importance of base model selection when employing PC prior. In addition, the $\operatorname{Uniform}\left(0,0.5\right)$ prior performs well in our simulation settings, and it is able to recover the true value except when $\ell = 0$ lies at the boundary. However, as discussed in Section 4.1.2, the $\operatorname{Uniform}\left(0,0.5\right)$ prior does not allocate sufficient density around both base models, and is therefore not recommended. It is worth nothing that almost all priors cannot recover the truth at the $\ell = 0$ boundary, except for  the $0.5\times\operatorname{Beta}\left(a,b\right)$ prior with $a < 1$, and the PCU prior with small $\alpha$ values (when $U$ is fixed). This is not surprising since the estimate cannot go below $0$ and it is almost impossible to recover the truth if the truth is right at the boundary.

\subsection{Wrapped Cauchy Distribution}

For $p\left(\rho\right)$ we consider the $\operatorname{Beta}\left(a,b\right)$ and the PC priors (Equation 3.22). The data are sampled with $\rho=0, 0.01, 0.5, 0.99$; The hyperparameter for $\operatorname{Beta}\left(a,b\right)$ prior has values of $a, b \in \left\{0.5, 1, 2, 5\right\}$; The scaling parameters for PC prior are set with $U=0.6$, and $\alpha = $0.01, 0.1, 0.3, 0.5, 0.7, 0.9, 0.99. The right-hand side plot in \autoref{fig:rhat_card_and_wc} shows that both priors achieve good convergence, while the PC prior has slightly lower $\hat{R}$ values when the true $\rho$ is close to either boundary.

\begin{figure}[!ht]
  \centering
    \includegraphics[width=\textwidth]{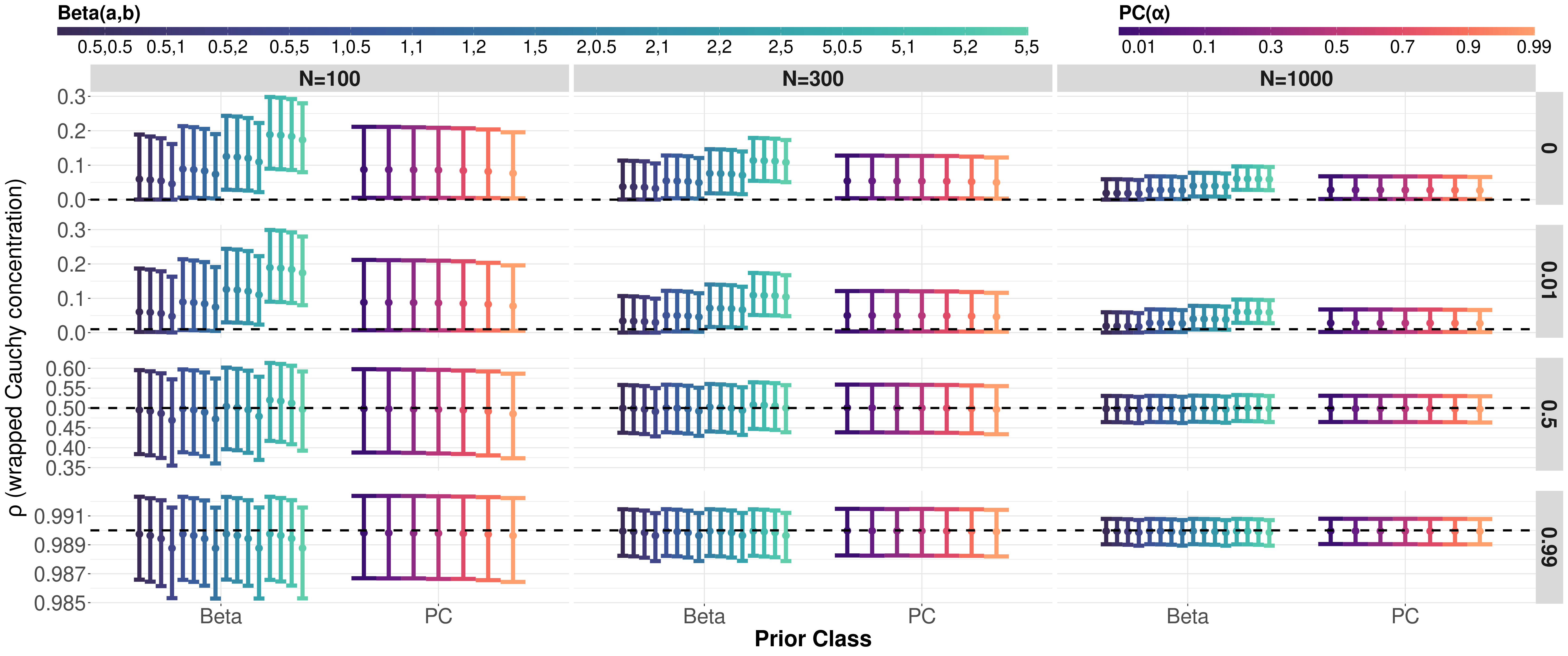}
    \caption{Simulation study for priors on the concentration parameter $\rho$ of the wrapped Cauchy distribution across different sample sizes and true $\rho$ values. The horizontal dashed line indicates the true $\rho$, with points showing posterior means and vertical bars indicating 95\% posterior credible intervals for each prior class.}
    \label{fig:wc_simulation}
\end{figure}

\autoref{fig:wc_simulation} shows that the PC priors demonstrate more stable and accurate performance than the $\operatorname{Beta}\left(a,b\right)$ prior. However, if sufficient prior information supports the use of a $\operatorname{Beta}\left(a,b\right)$ prior, as discussed in Section 4.1.3, we suggest that $\operatorname{Beta}\left(a,b\right)$ prior with $a < 1$ are appropriate choices. This behavior is also evident in \autoref{fig:wc_simulation}. For simulation setting with $\rho=0$, the $\operatorname{Beta}\left(a,b\right)$ prior with $a=0.5$ produces reasonable estimates, with the credible interval successfully recovering the boundary base model.

Through the comparative study presented in this section, we have illustrated a strategy for selecting priors that discourage unnecessarily complex model specifications in circular distributions, emphasizing that the PC prior is consistently an appropriate choice. We acknowledge that other priors may also perform adequately and may outperform the PC prior when sufficient prior information is available to specify them appropriately. The procedure outlined in the previous sections offers a systematic way to examine prior behavior on the distance scale, allowing practitioners to assess whether a chosen prior may favor model specifications that are more complex than warranted by the available information.

\section{Application} \label{sec:application}

\begin{figure}[!ht]
  \centering
  \begin{minipage}[b]{0.33\textwidth}
    \includegraphics[width=\textwidth]{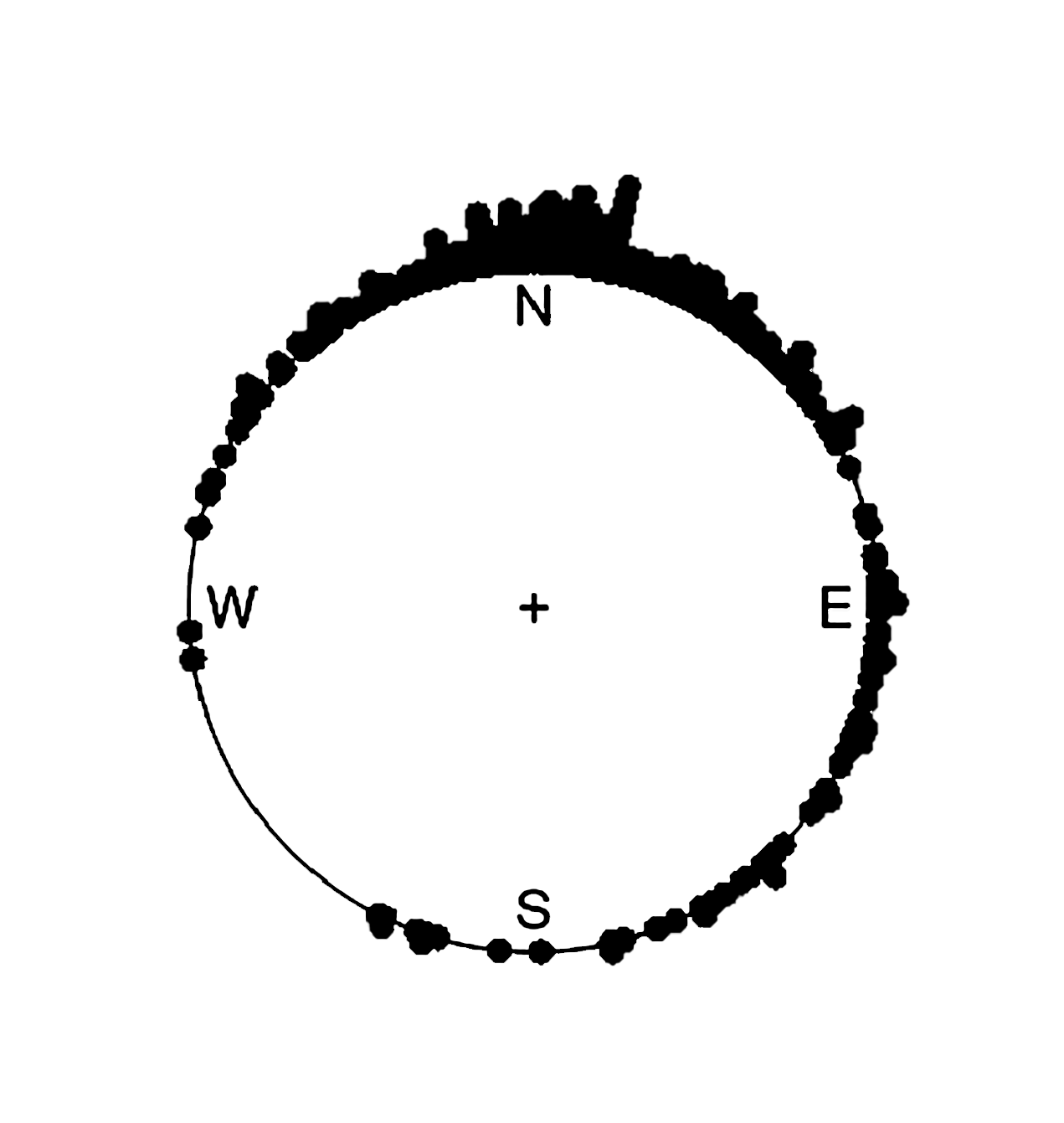}
    \caption{Wind data.}
    \label{fig:wind_data_point}
  \end{minipage}
  \hfill
  \begin{minipage}[b]{0.66\textwidth}
    \includegraphics[width=\textwidth]{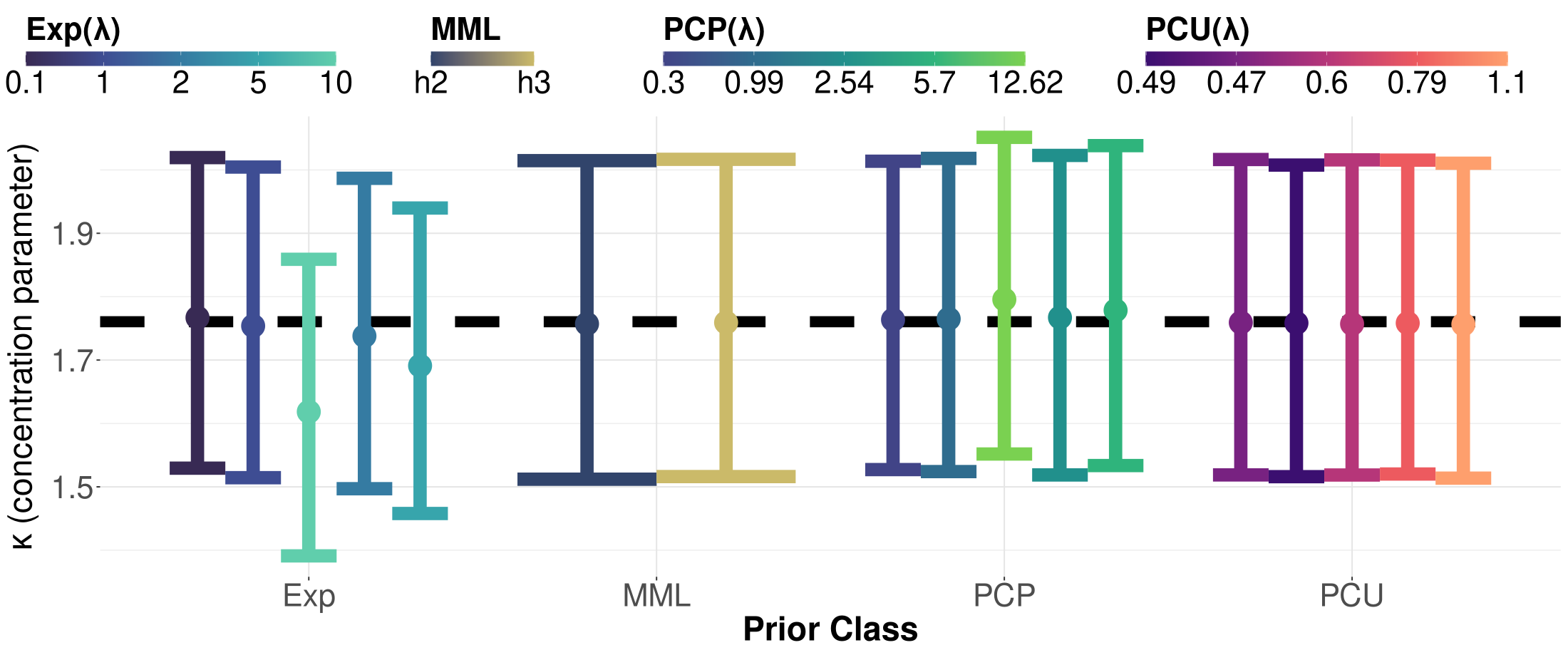}
    \caption{Posterior means and credible intervals of $\kappa$.}
    \label{fig:application_kappa_posterior}
  \end{minipage}
\end{figure}

In this section, we present a real-data case study to show the practical performance of the discussed priors. The data used in this study are the \texttt{wind} data (\autoref{fig:wind_data_point}) stored in "circular" package \citep{lund2017package} in R, recorded by a meteorological station in a place named "Col de la Roa" in the Italian Alps. The dataset contains 310 measures of daily wind direction from January 29th, 2001 to March 31st, 2001 covering the data span from 3 a.m. to 4 a.m. of the day. As a circular analogue of normal distribution, von Mises distribution is employed for our model, and the Bayesian model is the same as expressed in \autoref{sec:comparison_and_investigation_of_common_priors.simulation_study}. For comparison, the Stan model with $p\left(\kappa\right)$ being PC prior, $\operatorname{Gamma}\left(1,b\right)$ prior (exponential prior), $h_{2}$ prior and $h_{3}$ prior for $\kappa$ are constructed.

We choose $b=0.1, 1, 2, 5, 10$ for the $\operatorname{Gamma}\left(1,b\right)$ prior. From \autoref{fig:wind_data_point}, we can observe that this dataset is not highly concentrated. Therefore, for the PC prior, the circular uniform base model would be a more natural choice compared with the point mass base model. However, for the purpose of comparison and illustration, models with PC prior with both base models are constructed. We choose the scaling parameter $\lambda$ following the procedure discussed in \Cref{sec:a_principled_framework_for_prior_selection.model_complexity_check} with $U=\pi, \frac{\pi}{2}, \frac{\pi}{4}, \frac{\pi}{8}, \frac{\pi}{16}$, indicating different radians around $x=0$. The $\alpha$ value is chosen to be 0.2, 0.35, 0.5, 0.65, 0.8 respectively.

After fitting the model, the posterior mean and the 95\% credible interval for $\kappa$ are given in \autoref{fig:application_kappa_posterior}. In the plot, the horizontal dashed line indicates the frequentist maximum likelihood estimate of $\kappa$, with points showing posterior means and vertical bars indicating 95\% posterior credible intervals for each prior class. From the plot, the PC prior with circular uniform base model (PCU) and PC prior with point mass base model (PCP) fit the data well in terms of the stability of credible interval and posterior mean. For this dataset, as a more natural choice, the PCU prior is indeed robust than PCP prior, since the value of scaling parameter $\lambda$ does not vary much with the same given $U$ and $\alpha$ values. 

We evaluate predictive performance using the Expected Log Predictive Density (ELPD), Leave-One-Out Information Criterion (LOOIC) \citep{vehtari2017practical}, Watanabe–Akaike Information Criterion (WAIC) \citep{watanabe2010asymptotic}, and Deviance Information Criterion (DIC) \citep{spiegelhalter2002bayesian}. The ELPD, LOOIC and WAIC are computed through R package \texttt{loo} \citep{vehtari2015loo} version 2.9.0.

The ELPD evaluates out-of-sample predictive accuracy by summing the expected log predictive density for each observation. Under leave-one-out cross-validation, it is defined as
$$\text{ELPD}_{\text{loo}} = \sum_{i=1}^{N} \log p\left(y_{i} \mid y_{-i}\right),$$
where $p\left(y_{i} \mid y_{-i}\right)$ denotes the posterior predictive density of observation $i$ given the data with that observation excluded. Higher ELPD values indicate better predictive performance.

In the R package \texttt{loo}, $\text{ELPD}_{\text{loo}}$ is approximated using Pareto-smoothed importance sampling leave-one-out cross-validation (PSIS-LOO) \citep{vehtari2017practical, vehtari2024pareto}. The computation is based on the matrix of pointwise log-likelihood contributions evaluated at posterior draws. The reliability of the importance sampling approximation is assessed using Pareto-$k$ diagnostics. Following \citet{vehtari2017practical}, values $k \le 0.7$ indicate reliable approximation, values $0.7 < k \le 1$ suggest potential instability, and values $k > 1$ indicate that the approximation may fail and alternative approaches such as exact leave-one-out cross validation for those observations should be considered. We report the estimated ELPD together with its standard error ($\text{SE}\left(\text{ELPD}\right)$), as well as differences relative to the best-performing model ($\Delta\text{ELPD}$) and their associated standard errors ($\text{SE}\left(\Delta\text{ELPD}\right)$).

The LOOIC is defined as $\text{LOOIC} = -2\text{ELPD}_{\text{loo}}$, so that smaller values indicate better predictive performance. WAIC is computed from the same pointwise log-likelihood matrix and is defined as
$\text{WAIC} = -2 (\text{lppd} - p_{\text{WAIC}})$ where $$\text{lppd} = \sum_{i=1}^N \log \left( \frac{1}{S} \sum_{s=1}^S p\left(y_i \mid \theta^{\left(s\right)}\right) \right) \quad \text{and} \quad p_{\text{WAIC}} = \sum_{i=1}^N \text{Var}_{s=1}^S \left(\log p\left(y_i \mid \theta^{\left(s\right)}\right)\right).$$ We report LOOIC and WAIC together with its standard error and differences relative to the best-performing model and their associated standard errors. The DIC is difined as $\text{DIC} = \bar{D} + p_D$, where $D\left(\theta\right) = -2\log p\left(\mathbf{y}\mid \theta\right)$, $\bar{D}$ is its posterior mean, and $p_{D}$ denotes the effective number of parameters. Since DIC does not naturally provide a standard error, we report only differences in DIC ($\Delta\text{DIC}$) across models.

The predictive performance metrics are provided in Appendix C. Note that this comparison is intended to further illustrate the stability of the model, ensuring that the estimation does not deteriorate under different prior specifications. Therefore, we do not focus on identifying which prior yields the best predictive fit. For all models evaluated in this case study, all Pareto-$k$ values are $\le 0.7$, indicating that the importance sampling approximation is reliable and exact leave-one-out refitting is not required.

In this application, all models have similar performance with respect to the chosen predictive metrics. The $\Delta\text{ELPD}$, $\Delta\text{LOOIC}$, $\Delta\text{WAIC}$ and $\Delta\text{DIC}$ values suggest that the predictive performance under the PCP and PCU priors is more stable across different hyperparameter settings compared with the Exp prior. \autoref{fig:application_kappa_posterior} further shows that the posterior mean and credible interval under the Exp prior are quite sensitive to the choice of $\lambda$, and there is limited guidance for selecting the hyperparameters of the $\operatorname{Gamma}\left(1,b\right)$ prior. In addition, the $h_{2}$ and $h_{3}$ priors both perform well in this example. As discussed in \Cref{sec:comparison_and_investigation_of_common_priors.von_mises_distribution}, the $h_{2}$ prior also favors a circular uniform base model. Therefore, when the data have low concentration, the $h_{2}$ prior appears to be a reasonable choice for $\kappa$.

This practical example demonstrates that the PC priors have stable and comparable performance in circular scenarios, while ensuring that the simpler model is favored a priori.

\section{Conclusion and Outlook} \label{sec:conclusions_and_outlook}

A framework for selecting priors for parameters of circular distributions is essential for constructing robust Bayesian circular models and for avoiding unnecessarily complex specifications when the parameter space is difficult to interpret. The framework proposed in this paper focuses on assessing whether a prior adequately controls model complexity, thereby supporting a more reliable construction of Bayesian circular models. Additionally, it provides a simple and practical way to choose prior hyperparameters so that the resulting priors can be either strongly or weakly informative, depending on the level of prior knowledge available.

The derived PC priors are inherently designed to regularize model complexity and remain stable choices, particularly when prior knowledge is limited or uncertain. In the empirical studies considered in this paper, posterior sampling under the PC prior showed no evidence of mixing or numerical instability, including in boundary scenarios. It is worth noting that the advantage of the PC prior lies in its control of model complexity, which helps guard against poorly supported model specifications rather than aiming to achieve the best possible fit. However, when reliable and well-calibrated prior information is available, carefully specified informative priors tailored to a particular setting may provide improved performance. In these situations, the proposed framework can be employed to evaluate prior behavior with respect to model complexity and assess the suitability of alternative priors. The present analysis does not include a direct numerical diagnostic for excessive model complexity. Instead, model complexity is assessed theoretically by examining whether a prior places sufficient mass near the chosen base model, which can be clearly seen when representing the prior on the distance scale. While a dedicated quantitative metric could further support this assessment, introducing such tools is beyond the scope of this work and may be explored in future research.

Building upon the prior selection strategies and the PC prior for circular distributions addressed in this paper, we develop a general Bayesian regression framework capable of handling circular responses, circular covariates, linear covariates, or any combination of these \citep{ye2026bayesian}. Future efforts will also focus on extending the prior selection framework to accommodate joint priors and providing conditional PC priors, which are important for handling multi-parameter and multi-modal distributions and models.

\section*{Supplementary materials}
The supplementary materials include the appendix for the derivation of the PC priors, and the code for simulation studies (in ibex cluster) and application (in R). They are available at \url{https://github.com/XiangYEstats/PCpriors-circular}.

\bibliographystyle{abbrvnat}
\bibliography{references}

\end{document}